\documentclass[useAMS,usenatbib]{mnras}

\usepackage{amssymb}
\usepackage{amsmath}
\usepackage{amsfonts}
\usepackage{graphicx}
\usepackage{color}
\usepackage{url}
\usepackage[T1]{fontenc}
\def\aj{AJ}% Astronomical Journal

\def\apj{ApJ}% Astrophysical Journal
\def\apjl{ApJ}% Astrophysical Journal, Letters
\def\apjs{ApJS}% Astrophysical Journal, Supplement
\def\aap{A\&A}% Astronomy and Astrophysics
\def\mnras{MNRAS}% Monthly Notices of the RAS
\def\nat{Nature}% Nature
\def\FeH{\mathrm{[Fe/H]}}
\def\aFe{[\alpha/\mathrm{Fe}]}
\def\e{{\rm e}}
{\newif\ifnotend
\notendtrue
\def\veclist{ABCDEFGHIJKLMNOPQRSTUVWXYZabcdefghijklmnopqrstuvwxyz.}
\def\top#1#2.{#1}
\def\tail#1#2.{#2.}
\loop\expandafter\xdef\csname v\expandafter\top\veclist\endcsname%
{{\noexpand\bf\expandafter\top\veclist}}
\edef\veclist{\expandafter\tail\veclist}
\if\veclist.\notendfalse\fi\ifnotend\repeat}
\def\vtheta{\mathbf{\theta}}
\def\kpc{\,{\rm kpc}}
\def\Gyr{\,{\rm Gyr}}
\def\kms{\,\mathrm{km\,s}^{-1}}

\newcommand{\degree}{\ensuremath{^\circ}}
\newcommand{\diff}{\mathrm{d}}
\title[An EDF for halo K giants]{Characterising stellar halo populations I: An extended distribution function for halo K giants}
\author[Payel Das]{Payel Das $^{1}$\thanks{E-mail:
payel.das@physics.ox.ac.uk} and James Binney$^{1}$\\
$^{1}$Rudolf Peierls Centre for Theoretical Physics, University of Oxford, OX1 3NP, UK}

\begin{document}

\pagerange{\pageref{firstpage}--\pageref{lastpage}} \pubyear{2016}

\maketitle

\label{firstpage}

\begin{abstract}
We fit an Extended Distribution Function (EDF) to K giants in the Sloan
Extension for Galactic Understanding and Exploration (SEGUE) survey. These
stars are detected to radii $\sim80\kpc$ and span a wide range in [Fe/H]. Our
EDF, which depends on [Fe/H] in addition to actions, encodes the entanglement
of metallicity with dynamics within the Galaxy's stellar halo. Our
maximum-likelihood fit of the EDF to the data allows us to model the
survey's selection function.

The density profile of the K giants steepens with radius from a slope
$\sim-2$ to $\sim-4$ at large radii. The halo's axis ratio increases with
radius from 0.7 to almost unity. The metal-rich stars are more tightly
confined in action space than the metal-poor stars and form a more flattened
structure. A weak metallicity gradient $\sim-0.001\,$dex/kpc,
a small gradient in the dispersion in [Fe/H] of $\sim0.001\,$dex/kpc, and a
higher degree of radial anistropy in metal-richer stars result.  Lognormal
components with peaks at $\sim-1.5$ and $\sim-2.3$ are required to capture
the overall metallicity distribution, suggestive of the existence of two
populations of K giants.

The spherical anisotropy parameter varies between 0.3 in the inner halo to
isotropic in the outer halo. If the Sagittarius stream is included, a very similar model is found but with a stronger degree of radial anisotropy throughout.
\end{abstract}


\section[]{Introduction}
% Introduction to stellar halo as a collisionless system
About 1$\%$ of the stellar mass in the Galaxy comprises the stellar halo,
primarily in the form of old and metal-poor stars. Some of these stars may
have formed in-situ, but others come from disrupted satellite
galaxies and globular clusters
\citep{ibata+95,majewski+03,belokurov+07,bullock+05,abadi+06,cooper+11}. An
essential step towards  distinguishing between processes for halo formation
is to gain  a clear characterisation of the halo's present chemodynamical
structure. 

Since the Galaxy's potential is in an approximately steady state, we can
construct steady-state models of the halo by adopting distribution functions
(DFs) $f(\vJ)$ that depend only on the constants of stellar motion $J_i$
\citep{jeans+16}. There are clear advantages in taking these constants of
motion to be the actions. First, action coordinates can be complemented by
canonically conjugate variables, the angles, to obtain a complete coordinate
system for phase space. Second, it is straightforward to add distinct DFs
$f(\vJ)$ for the thin and thick discs, the bulge and the dark halo to the DF
$f(\vJ)$ of the stellar halo to build up a complete Galaxy model
\citep{binney_piffl2015}. Third, actions are adiabatic invariants and
therefore can be used to examine phenomena such as adiabatic contraction
\citep[e.g.][]{piffl+15}. Finally, the actions $J_r$, $J_\phi$ and $J_z$ have
simple physical interpretations in that they approximately describe
excursions of an orbit in the radial, azimuthal and vertical directions.

A DF $f(\vJ)$ maps the halo into a distribution of stars in
an easily imagined  three-dimensional action space. In general we expect the dynamics of a
population of stars to depend on their chemistry, so we assume that $f$ is a
function of [Fe/H] in addition to  $\vJ$. That is, we assume that the halo can
be well represented by an Extended Distribution Function (EDF) of the type
introduced by \cite{sanders+15} for the Galaxy's disc(s). 

From an EDF one can predict the velocity distribution of stars of any metallicity
at any location, so an EDF provides the power to pull together results from
several surveys into a single coherent picture.

\cite{bell+08} established that much of the halo is contributed by
substructures that are thought to be relics of disrupted satellites and
globular clusters. An EDF of the form $f(\vJ,\FeH)$ can represent such
structures only to the extent that they are old enough to have become well
phase mixed. Moreover, each disruption event will have created a clump of
stars in action space, so the halo's EDF is likely not a smooth function of
$\vJ$. Nonetheless, in this paper we investigate the extent to which existing
data for the halo can be represented by an EDF which {\it is\/} a smooth
function of $\vJ$. If this proves possible, it will suggest that the data do
not yet provide sufficient phase-space resolution to demonstrate the existence of clumps
in action space. Alternatively, an unsuccessful attempt to fit a smooth
EDF to the data will bring into focus action-space structures arising from
accretion events. 

% The perspective from surveys
The stellar halo can be traced through several stellar populations. Blue
horizontal-branch (BHB) stars are luminous, primarily metal-poor, and their
distances can be related simply to their apparent magnitudes, colours, and
metallicities \citep{deason+11,fermani+13a}. RR Lyrae stars are also
primarily metal-poor and are associated with a period-luminosity function
that can be used to determine their distances.  Metal-poor K giants, which have  high
luminosities, provide useful tracers of the outer halo.

%% Density and flattening of the stellar halo
Early efforts to determine the halo's density profile were limited to
Galactocentric radii $r\sim 20 - 30\kpc$. The halo proved to be oblate with an
axis ratio $q\sim 0.5 - 0.8$, and its density declined as $r^{-\alpha}$
with $\alpha\sim 2 - 4$
\citep[e.g.][]{preston+91,robin+00,yanny+00,juric+08}.  More recent work,
reaching to greater distances, has found evidence for a break in the stellar
density profile at $r\sim20 - 30$ kpc. Star counts of main-sequence turn-off
stars \citep{bell+08,sesar+11}, RR Lyrae stars \citep{watkins+09,sesar+13},
BHB stars \citep{deason+11}, and K giants
\citep{kafle+13} all exhibit a break in the number density around  $R =15 -
25\kpc$ with $\alpha\sim2 - 3$ at $r<25\kpc$ and $\alpha\sim 3.8 - 5$ further
out. An examination of data for BHBs and blue stragglers from the SDSS by
\cite{deason+14} found evidence that $\alpha\sim6$ beyond $50\kpc$ and
$\alpha\sim6 - 10$ at still larger radii. More recently however,
\cite{xue+15} found that if the halo's flattening is permitted to decrease
with radius in the case of K giants observed in SEGUE, the
best-fitting model has $\alpha=4.2\pm0.1$ at all radii, but a break in the
flattening at $\sim 20\kpc$ from axis ratio $q=0.5 - 0.6$ at $r<20\kpc$ to
$q\sim0.8$ much further out.

%% Dynamics
\cite{gould+03} and \cite{smith+09} find that the
velocity ellipsoid is generally well aligned with spherical polar
coordinates, as it would be if the Galaxy's potential were spherical.

It is unclear whether the halo rotates \citep{fermani+13a}, and if so,
whether at the largest radii it rotates in the opposite sense to the disc
\citep{gould+03,beers+12,schonrich+11,schonrich+14}. Dynamical models in a
spherical potential have
generally found radial velocity dispersions to be bigger than the tangential
dispersions \citep{deason+12a,kafle+14,williams+15b}. Some authors
have, however, found the opposite in the region $15<r<25$ kpc
\citep{kafle+12,deason+13}. If there really is a radial range in which
tangential velocity dispersion is dominant, this may reflect the presence of
substructures in the halo, especially the Sagittarius stream
\citep{chiba+00,king+15}.  

%% Metallicity and connection to phase-space
The existence of a metallicity gradient in the halo is controversial. Early
work such as that of \cite{zinn+85} found no gradient outside the solar
radius, but \cite{carollo+07} claim a  negative gradient, with the
outer halo significantly more metal poor than the local halo. This was later confirmed by several studies \citep[e.g.][]{dejong+10,kafle+13,allende+14,chen+14}, with some also finding metal-poorer stars in retrograde motion and metal-richer stars in prograde motion \citep{carollo+07,kafle+13}. \cite{schonrich+14} argue that metal-poorer stars can be seen at greater distances than metal-richer stars, and if this effect is not correctly included in the adopted selection function, a
metallicity gradient can be erroneously inferred. However work by \cite{peng+12} examining $\sim40\,000$ main-sequence stars found little
evidence for a gradient, although their analysis is limited to metallicities above -2.0. The latest study on the
topic by \cite{xue+15}, who take the selection function into account in a
sample of K giant stars, does find a modest, but significant metallicity
gradient within the outer halo.

% Importance and need for constructing a chemodynamical model
To obtain a coherent picture of the current state of the halo, a model needs
to be created that simultaneously considers the various constraints on the
density, the kinematics, and the chemistry, with a detailed consideration of
the various selection effects in play. A forward-modelling approach based on
an EDF is best because then one can take full account of the constraints
imposed by dynamics and one can model the impact of the selection function on
the data. In this paper, we develop an EDF for K giants, which span a range
of metallicities.
\section[]{A model stellar halo}
An extended distribution function (EDF) gives the probability density of
stars in the space spanned by the phase-space coordinates
$(\mathbf{x},\mathbf{v})$ and the variables that characterise stars, such as
mass $m$, age $\tau$, and chemistry ([Fe/H], $\aFe,\ldots$). Given the
success of the hypothesis of a universal initial mass function (IMF) in
astronomy, we assume that the halo's EDF is simply proportional to this IMF
below the mass at which the stellar lifetime becomes equal to $\tau$, and
zero at higher masses.  We assume that all halo stars have age $\tau=11\Gyr$
\citep{jofre+11}, and adopt the IMF of \cite{kroupa+93}
\begin{equation}
	\epsilon(m) = 
	\begin{cases}
		0.035m^{-1.5} &\mathrm{if} \, 0.08 \leq m < 0.5\\
		0.019m^{-2.2} &\mathrm{if} \, 0.5 \leq m < 1.0\\
		0.019m^{-2.7} &\mathrm{if} \, m \geq 1.0\,.
	\end{cases}
\end{equation}
We use the $\alpha$-enhanced B.A.S.T.I. isochrones of \cite{pietrin+06}. With
these simplifying assumptions, the only explicit dependencies of the EDF are
on phase-space coordinates and [Fe/H], so the EDF can be considered either a
function $f(\mathbf{x},\mathbf{v},\FeH)$ or a function $f(\mathbf{J},\FeH)$.

We take the gravitational potential to be axisymmetric, so the natural choice
of phase-space coordinates are cylindrical polar coordinates
$(R,\phi,z,v_R,v_{\phi},v_z)$. For actions we use $J_r, J_\phi\equiv L_z$ and
$J_z$.  We use the St\"{a}ckel Fudge \citep{binney+12} to convert
$(\vx,\vv)$ to $\vJ$.

We adopt a left-handed coordinate system in which positive $v_R$ is away from
the Galactic centre and positive $v_{\phi}$ is in the direction of Galactic
rotation. To convert from the Galactocentric coordinates to the heliocentric
coordinates we assume that the Sun is located at $(R_0,z_0) =
(8.3,0.014)\kpc$ \citep{schonrich+12,binney+97}, that the Local Standard of
Rest (LSR) has an azimuthal velocity of $238\kms$, and that the velocity of
the Sun relative to the LSR is $(v_R,v_{\phi},v_z) = (-14.0, 12.24,
7.25)\kms$ \citep{schonrich+12}.
\subsection[]{An extended distribution function (EDF)}
\begin{figure}
	\centering
	\includegraphics[scale=0.5]{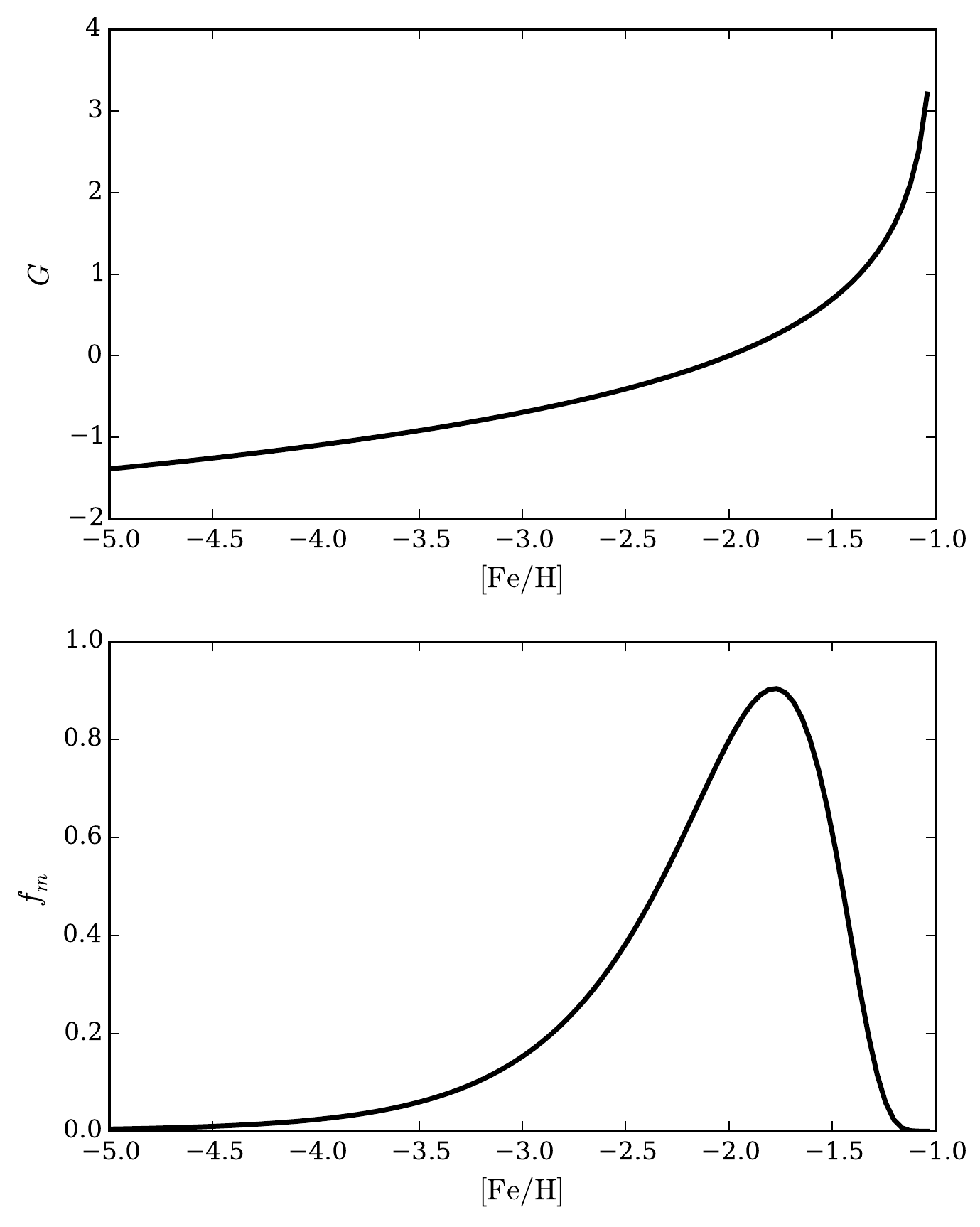}
	\caption{Top: the relation between [Fe/H] and the variable  $G$
	for $\FeH_{\mathrm{max}} = -1.0$. Bottom: the overall metallicity
distribution function $f_\mathrm{m}(\FeH)$ when $\FeH_{\mathrm{max}} = -1.0$ and $\sigma=0.5$.\label{fig:G_mdf}}
\end{figure}

The formulae below are significantly simplified by introducing as a measure
of metallicity the variable
\begin{equation}\label{eq:defsG}
G \equiv -\ln (\FeH_{\mathrm{max}} - \FeH).
\end{equation}
$G$ is an increasing function of [Fe/H] with a vertical
asymptote at $\FeH_{\mathrm{max}}$.
The upper panel of Fig.~\ref{fig:G_mdf} shows the relationship between $G$
and [Fe/H] when $\FeH_{\mathrm{max}} = -1.0$.

Our EDF is 
\begin{equation}\label{eq:EDF}
{\mathrm{d}N\over\mathrm{d}^3\mathbf{x}\,\mathrm{d}^3\mathbf{v}\,\mathrm{d}\FeH}
=	f(\mathbf{J},G) = Af_\mathrm{m}(G)f_{\mathrm{ps}}(\mathbf{J},G),
\end{equation}
where $A$ is a normalisation constant enforcing a total probability of one.
For $f_{\mathrm{m}}(G)$, which is approximately the overall metallicity DF,
we adopt 
\begin{equation}\label{eq:MDF}
{\mathrm{d}N\over\mathrm{d}\FeH}
=	f_{\mathrm{m}}(G) =
	\e^{G}\frac{\e^{-\frac{G^2}{2\sigma^2}}}{\sigma\sqrt{2\pi}}.
\end{equation}
This function, which is plotted in the lower panel of Fig.~\ref{fig:G_mdf}
for $\FeH_{\mathrm{max}} = -1.0$ and $\sigma=0.5$, provides a well defined
peak at $G=\sigma^2 \Rightarrow \FeH=-1.78$ and a tail towards lower
metallicities. Although ${\mathrm{d}N/\mathrm{d}G}$ is a Gaussian in $G$, ${\mathrm{d}N/\mathrm{d}\FeH}$ is not a Gaussian in either $G$ or [Fe/H]. Several of these components could be superimposed to capture
the overal metallicity DF; we start with just one.

The part of the EDF \eqref{eq:EDF} that involves $\vJ$,
$f_{\mathrm{ps}}(\mathbf{J},G)$, is a generalisation of the DF proposed by
\cite{posti+15} for spheroidal objects.  Specifically,
\begin{equation}\label{eq:postiDF}
	\begin{split}
		f_{\mathrm{ps}}(\mathbf{J},G)  &= \frac{[1+J_0/h(\mathbf{J})]^{\alpha(G)}}{[1+g(\mathbf{J})/J_0]^{\beta(G)}}\\
		h(\mathbf{J}) 				&= J_{\mathrm{core}} + a_rJ_r + a_{\phi}|J_{\phi}| + a_zJ_z\\
		g(\mathbf{J})  				&= b_rJ_r + b_{\phi}|J_{\phi}| + b_zJ_z\\
		\alpha(G) 		&= \alpha^{\prime}_0+\alpha_\mathrm{m}\e^{-G}\\
		\beta(G)  		&= \beta^{\prime}_0+\beta_\mathrm{m}\e^{-G},
	\end{split}
\end{equation}
For $|\vJ|\gg J_0$, $f_{\rm ps}$ is dominated by $g(\vJ)$, which is a
homogeneous function of the actions, and for $|\vJ|\ll J_0$, $f_{\rm ps}$ is
dominated by $h(\mathbf{J})$, which would also be a homogeneous function of
$\vJ$ but for a dependence on the core action, $J_{\mathrm{core}}$. This
dependence ensures that $f_{\rm ps}$ tends to a finite value as $\vJ\to0$
(the orbit that involves sitting at the centre of the Galaxy). Outside the
tiny core region, $h(\mathbf{J})$ ensures that the phase-space density
declines as $|\vJ|^{-\alpha}$ for $|\vJ|<J_0$, while $g(\vJ)$ ensures that
the phase-space density declines as $|\vJ|^{-\beta}$ at much larger actions.

\cite{posti+15} showed that DFs of the form \eqref{eq:postiDF} generate
self-gravitating systems in which the density profile approximates a double
power law in radius, so $\rho\propto r^{-\gamma}$, where the inner and outer
values of $\gamma$ are related in a simple way to $\alpha$, and $\beta$,
respectively. Thus by making $\alpha$ and $\beta$ functions of metallicity we
can arrange for the metal-rich stars (for example) to form a more compact
body than the metal-poor stars, with the consequence that there is a
metallicity gradient within the halo.

Recasting the metallicity dependence of $\alpha,\beta$ in terms of
$[\mathrm{Fe/H}]$ rather than $G$, gives 
\begin{equation}
	\begin{split}
	\alpha &= \alpha_0-\alpha_\mathrm{m}[\mathrm{Fe/H}]\\
	\beta &= \beta_0-\beta_\mathrm{m}[\mathrm{Fe/H}]\\
	\textrm{where}&\\
	\alpha_0 &= \alpha^{\prime}_0 + \alpha_\mathrm{m}[\mathrm{Fe/H}]_{\mathrm{max}}\\
	\alpha_0 &= \alpha^{\prime}_0 + \alpha_\mathrm{m}[\mathrm{Fe/H}]_{\mathrm{max}}.
	\end{split}
\end{equation}

 Since $f_{\rm ps}$ is an even function of $J_\phi$, it does not endow the
halo with rotation.

The parameters $(a_r,a_{\phi},a_z,b_r,b_{\phi},b_z)$ control the shape of the
density distribution and the velocity ellipsoids \citep{posti+15}. Rescaling all
the $a_i$ and $b_i$ by the same factor will have no effect on the model if it
is accompanied by rescaling of $J_{\rm core}$ and $J_0$. We eliminate this
degeneracy by imposing the conditions $\sum_i a_i=\sum_i b_i=3$. 

Reducing $a_i$ or $b_i$ increases the velocity dispersion in the $i$th
coordinate direction in the inner/outer halo. Increasing $\sigma_z$ reduces
the flattening of the density distribution. The flattening can also be reduced
by lowering $\sigma_\phi$. Thus the $a_i$ and $b_i$ control both the
kinematics and the physical shape of the halo. We could have made them
functions of [Fe/H], but in this first attempt at an EDF for the halo we do
not do so.

\begin{table}
 \centering
  \caption{Parameters of the Galactic potential.\label{tab:potpars}}
  \begin{tabular}{lll}
  	\hline
  	Component     		&Parameter     					 &Value\\
  	\hline
  	Thin          		&$R_\mathrm{d}$ (kpc)     				 &2.682\\
  					    &$z_\mathrm{d}$ (kpc)     				 &0.196\\
  					    &$\Sigma_\mathrm{d} (M_{\odot}$kpc$^{-2}$)  &5.707$\times10^8$\\
  	\hline
  	Thick       			&$R_\mathrm{d}$ (kpc)     				 &2.682\\
  					    &$z_\mathrm{d}$ (kpc)     				 &0.701\\
  					    &$\Sigma_\mathrm{d} (M_{\odot}$kpc$^{-2}$)  &2.510$\times10^8$\\
  	\hline
  	Gas         			&$R_\mathrm{d}$ (kpc)     				 &5.365\\
  					    &$z_\mathrm{d}$ (kpc)     				 &0.040\\
  					    &$\Sigma_\mathrm{d} (M_{\odot}$kpc$^{-2}$)  &9.451$\times10^7$\\
  					    &$R_{hole}$ (kpc)				 &4.000\\
  	\hline
  	Bulge	  			&$\rho_0 (M_{\odot}$kpc$^{-3}$)    &9.490$\times10^{10}$\\
  						&$q$								 &0.500\\
  						&$\gamma$						 &0.000\\
  						&$\delta$						 &1.800\\
  						&$r_0$ (kpc)						 &0.075\\
  						&$r_\mathrm{t}$ (kpc)					     &2.100\\
  	\hline
  	Dark halo    &$\rho_0 (M_{\odot}$kpc$^{-3}$)    &1.815$\times10^7$\\
  						&$q$								 &1.000\\
  						&$\gamma$						 &1.000\\
  						&$\delta$						 &3.000\\
  						&$r_0$ (kpc)						 &14.434\\
  						&$r_\mathrm{t}$ (kpc)
						&$\infty$\\
  	\hline
  \end{tabular}
\end{table}
\subsection[]{The gravitational potential}
Our potential is based on the functional forms proposed by \cite{dehnen+98}.
It is generated by thin and thick stellar discs, a gas disc, and two spheroids
representing the bulge and the dark halo. The densities of the discs are
given by
\begin{equation}
	\rho_\mathrm{d}(R,z) = \frac{\Sigma_0}{2z_\mathrm{d}}\exp\left[-\left(\frac{R}{R_\mathrm{d}} + \frac{|z|}{z_\mathrm{d}} + \frac{R_{\mathrm{hole}}}{R}\right) \right],
\end{equation}
where $R_\mathrm{d}$ is the scale length, $z_\mathrm{d}$, is the scale height, and
$R_{\mathrm{hole}}$ controls the size of the hole at the centre of the disc,
which is only non-zero for the gas disc. The densities of the bulge and dark
halo are given by
\begin{equation}
	\begin{split}
		\rho(R,z) = & \rho_0\frac{(1+m)^{(\gamma-\delta)}}{m^{\gamma}}\exp\left[-(mr_0/r_{\mathrm{t}})^2\right]\\
		\mathrm{where\ } m(R,z) =& \sqrt{(R/r_0)^2 + (z/qr_0)^2}.
	\end{split}
\end{equation}
Here $\rho_0$ sets the density scale, $r_0$ is a scale radius, and the
parameter $q$ is the axis ratio of the isodensity surfaces. The exponents
$\gamma$ and $\delta$ control the inner and outer slopes of the radial
density profile, and $r_\mathrm{t}$ is a truncation radius. 

 The adopted parameter values are taken from \cite{piffl+14} and given in Table \ref{tab:potpars}. The dark
halo is assumed to be a spherical NFW halo, and not truncated
($r_\mathrm{t}=\infty$). The contribution of the stellar halo, which has
negligible mass, can be considered included in the contributions of the bulge
and dark halo.

\section[]{Observational constraints}
\begin{figure*}
	\centering
	\includegraphics[scale=0.6]{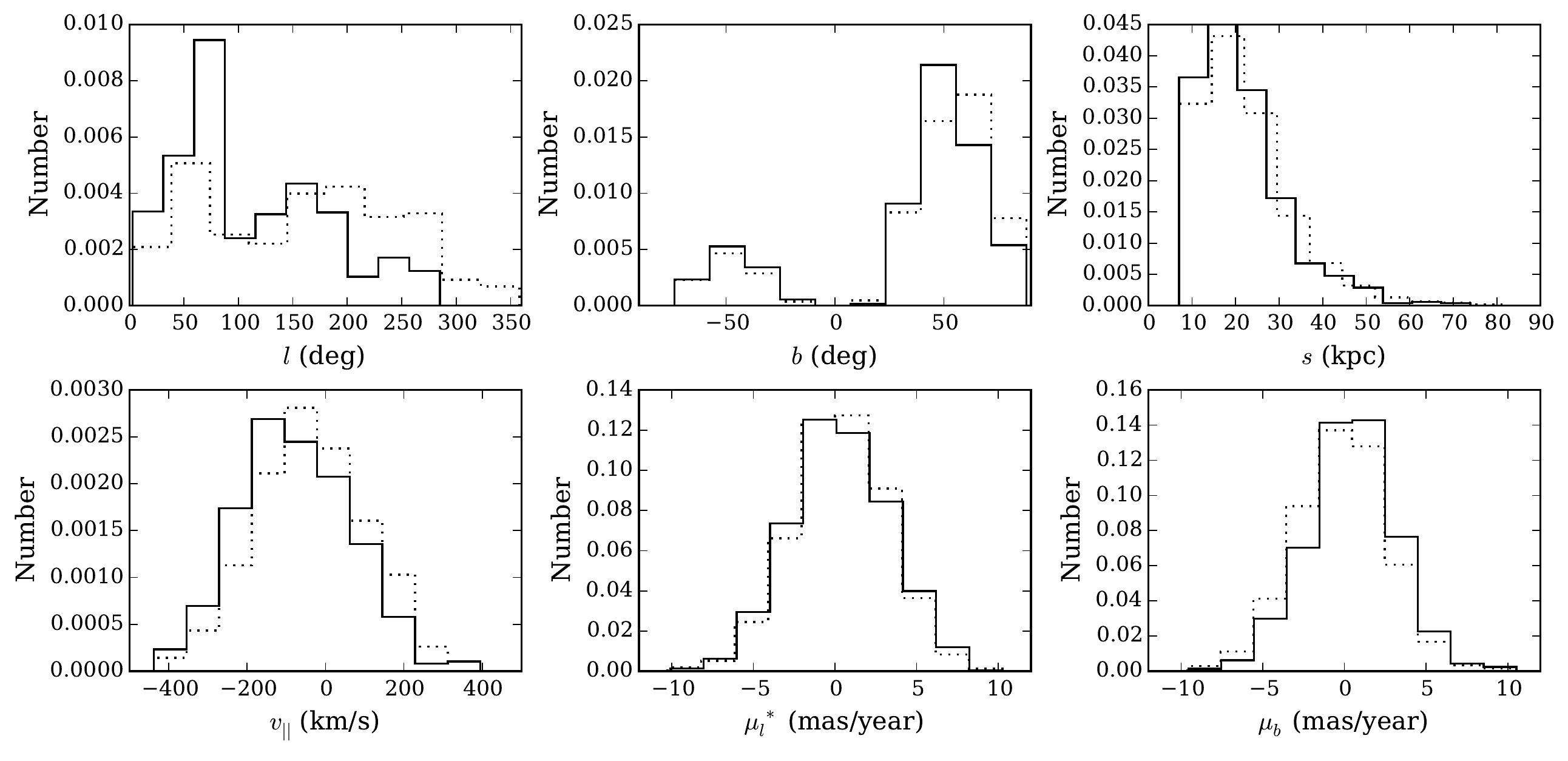}
	\caption{One-dimensional distributions in Galactic coordinates of phase-space observables with (dotted) and without (solid) the Sagittarius stream.\label{fig:phasespacehist}}
\end{figure*}
Here, we introduce the observations that we will use to constrain parameters
of the EDF. An ideal catalogue would provide Galactic coordinates $(l,b)$,
parallaxes $\varpi$, proper motions $(\mu_l^*=\dot l\cos b, \mu_b=\dot b)$,
apparent magnitudes, and colours, from which a photometric metallicity can be
derived $\FeH_{\rm ph}$. If spectra are also available, we have additionally
line-of-sight velocities $v_{||}$, spectroscopic metallicities $\FeH$,
$\alpha$-abundances, effective temperatures $T_{\mathrm{eff}}$, and surface
gravities $\log g$.  From isochrones we can relate apparent magnitudes to a
heliocentric distance\footnote{We shorten `heliocentric distance' to
distance for the remainder of the paper.} $s$.

\subsection[]{SEGUE data}

We take values for the seven components of the vector
\begin{equation}\label{eqn:obs}
	\mathbf{u} = (l,b,s,v_{||},\mu_l^*,\mu_b,\FeH),
\end{equation}
from Data Release 9 of the SDSS. These data form part of the Sloan Extension
for Galactic Understanding and Exploration survey II (SEGUE II). We do not
use SEGUE-I data because the selection function of these data is hard to
determine. We use only the K-giant sample of \cite{xue+14}. We supplement the catalogue values with proper motions downloaded from SkyServer's CasJobs\footnote{\url{http://skyserver.sdss.org/CasJobs/}}, by cross matching the data sets with the requirement that right ascensions
and declinations match within 15 arcsec. We remove stars on cluster and test
plates and apply the cut $\FeH\le-1.4$ to reduce the contamination from
disc stars \citep{schonrich+12}. We compile two data sets, one including the Sagittarius stream
and one that does not include plates that intersect the two polygons given
by \cite{fermani+13a} as containing the stream.  The final data sets contain
1689 and 1096 stars.
\begin{table*}
 \centering
  \caption{Combined selection criteria for SEGUE-II K giants according to `star type' (KG are K giants, RKG are red K giants, IKG are I-colour K giants, and PMKG are proper motion K giants) in terms of the proper motions, apparent magnitude, colour indices, and metallicity. $\mu$ is the total proper motion, $u$, $g$, $r$, and $i$ refer to apparent magnitudes in Sloan's $ugriz$ colour-magnitude system, and the I-colour is a photometric metallicity indicator for stars in the colour range $0.5 <(g-r) < 0.8$ given by I-colour$=-0.436u + 1.129g - 0.574i + 0.1984$ \citep{lenz+98}. A * indicates the modelled sample. The number in brackets gives the number of stars after excluding the Sagittarius stream. \label{tab:selfunc}}
  \begin{tabular}{llllllll}
  	\hline
	 $\#$ 	 	 &Star type   &Proper motions	&Apparent  	    &Colour				     &Metallicity\\	
		 		 &	 		 &				&magnitude 		&					     &\\			
		 		 &			 & (mas/yr)		& (mag)			& (mag)					 & (dex)\\
	\hline
	31 (12)		 &RKG		  &$\mu<11$ 		&$15.5<g<18.5$  &$0.8<(g-r)< 1.3$	     &$\FeH\le-1.4$\\
				&			 &			    &$r>15.0$       &$(u-g)>0.84(g-r)+1.758$  &\\  
				&		 	 &              & 				&$(u-g)<2.4(g-r)+0.73$   &\\  
	\hline	
	*1689 (1096)	 &IKG		  &$\mu<11$	 		&$15.5<g<18.5$	&$0.7<(u-g)<3$			 &$\FeH\le-1.4$\\
				&		 	 &			  		&$r>15.0$        &$0.5<(g-r)<0.8$		 &\\  	 
				&		 	 &		      &                 		&$0.1<(r-i)<0.6$		     &\\  
				&		 	 &			  &                 		&I-color$>0.09$		     &\\  
	\hline	
	105	 (75)	 &PMKG 		  &$\mu<7$	 		&$15.5<g<18.5$	&$0.8<(u-g)<1.2$		     &$\FeH\le-1.4$\\
				&		 	 &			  		&$r>15.0$       &$(u-g)>2.375(g-r)-0.45$  &\\  
				&		 	 &			        &                 &$(u-g)<0.84(g-r)+1.758$  &\\  
	\hline	
	\end{tabular}
\end{table*}
\subsection[]{Selection criteria}
Our selection functions are intersections of the original SEGUE-II
targeting criteria, and further criteria
imposed by \cite{xue+14} and by us.

\begin{figure}
	\centering
	\includegraphics[scale=0.6]{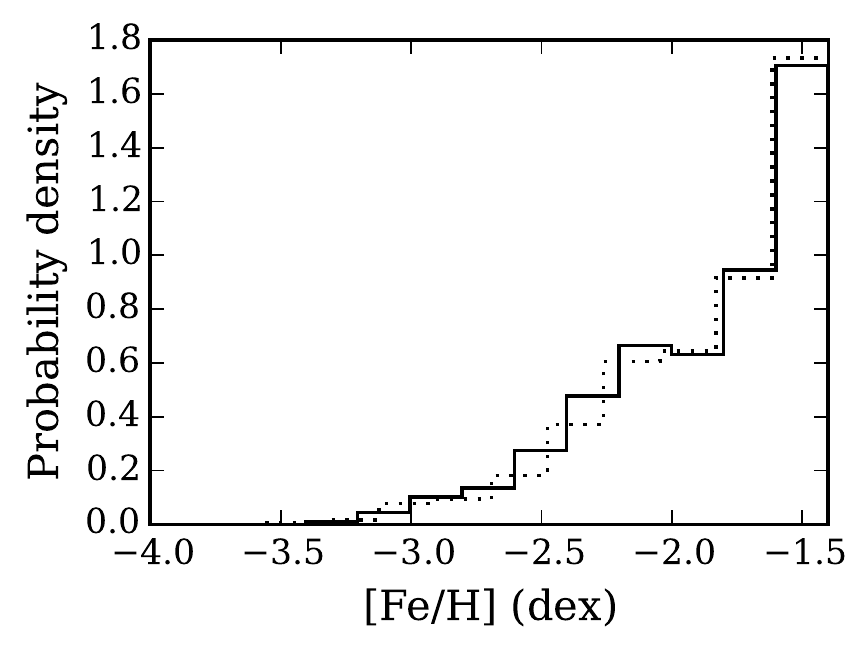}
	\caption{One-dimensional distribution of
	observed metallicities with (dotted) and without (solid) the Sagittarius stream.\label{fig:fehhist}}
\end{figure}

The selection on sky positions is given by the coverage of the SEGUE-II
plates, and on each plate by the completeness of the spectroscopic sample,
i.e., the fraction of available targets that were observed. We assume that
the completeness is independent of apparent magnitude, an assumption which we
checked for a handful of plates. The remaining selection criteria, relating
to proper motions, apparent magnitudes, colours, and metallicity are
summarised in Table \ref{tab:selfunc}. Grouping by selection criteria results
in three subsamples for the K giants.  The SEGUE-II I-colour K giants
comprises the largest sample, so we model these stars.  The phase-space and
metallicity distributions of the sample are shown in
Figs~\ref{fig:phasespacehist} and \ref{fig:fehhist}, with and without the
Sagittarius stream. Removing the Sagittarius stream affects the distributions
of the stars in sky positions, results in a slightly steeper observed density
radial profile, and shifts the distribution of $v_\parallel$ towards more
negative values because the Sagittarius stream has a mean
 $v_\parallel\sim150\kms$ \citep{ibata+97}.

\section[]{Fitting the data}
We determine the best-fit parameters of the EDF using the likelihood method
of \cite{mcmillan+13}, which \cite{sanders+15} used  to fit an EDF to
the Galaxy's discs. We now discuss the form of the likelihood and the methods
used to determine the contributing terms. 
\begin{figure}
\centering
\includegraphics[scale=0.5]{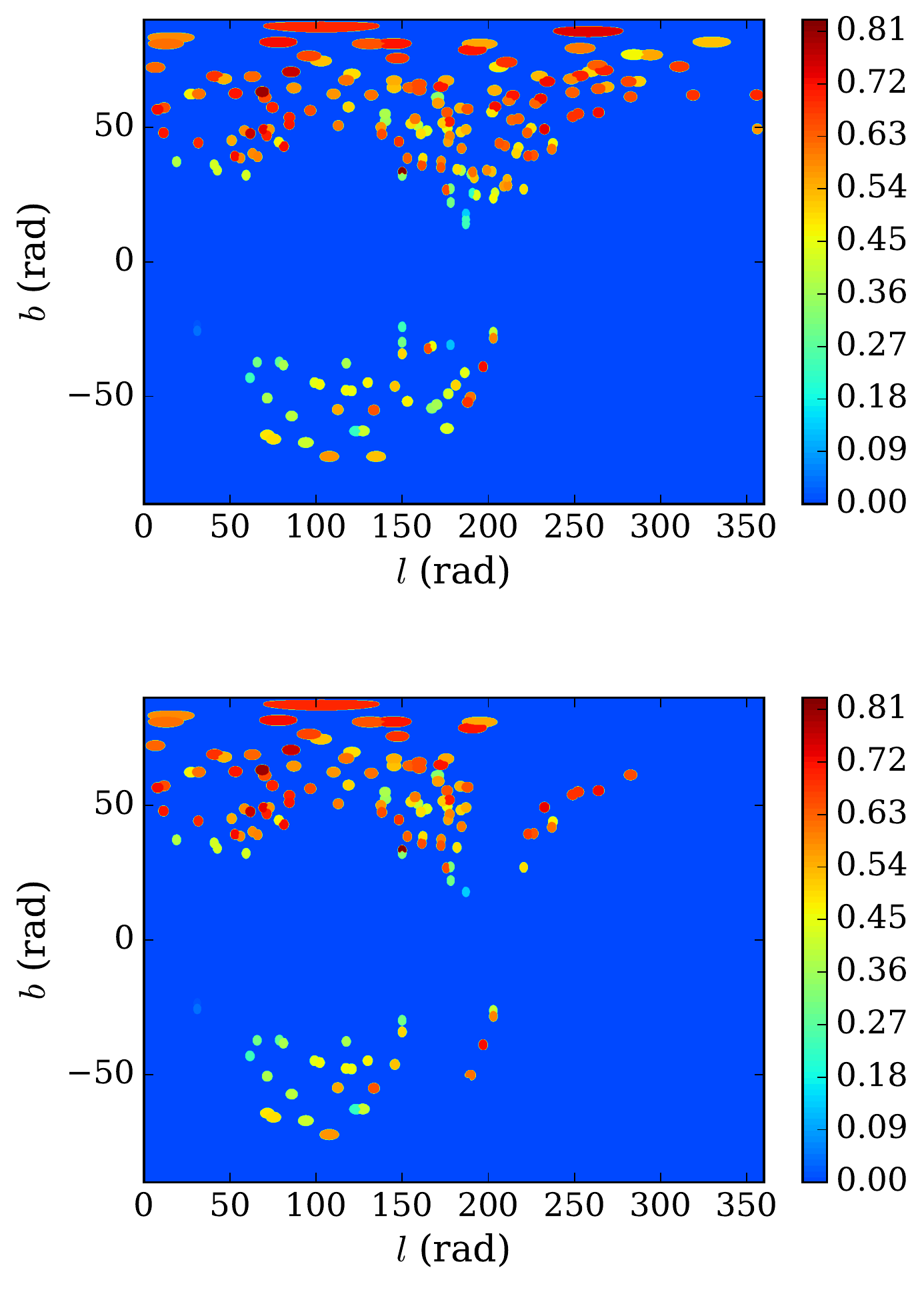}
 \caption{The colour scale shows as a function of sky positions
the selection probability $p(S|l,b)$ with (top) and without (bottom) the
Sagittarius stream included.\label{fig:lbsfmaps}}
\end{figure}
\subsection[]{The likelihood from Bayes' law}
The likelihood $\mathcal{L}$ of a model $M$ is given by the product over all
stars $i$ of the probabilities $\mathcal{L}^i=P(\mathbf{u}^i|SM)$. The $\mathcal{L}^i$ are the probabilities of measuring the star's catalogued coordinates $\mathbf{u}^i$ given the model $M$ and
that it is in the survey $S$. By Baye's law this is
\begin{equation}	
		\mathcal{L} = \prod_{k=i}^{n_*}\mathcal{L}^i 
= \prod_{k=i}^{n_*}\frac{P(S|\mathbf{u}^i)P(\mathbf{u}^i|M)}{P(S|M)},
\end{equation}
where $n_*$ is the number of stars. $P(S|\mathbf{u}^i)$ is the probability
that the star is in the survey given the observables i.e. the `selection
function'. $P(\mathbf{u}^i|M)$ is the EDF convolved with the error
distribution of the observables. $P(S|M)$ is 
the probability that a randomly chosen star in the model enters the
catalogue.  We maximise the total log-likelihood
\begin{multline}\label{eqn:logL}
	\log \mathcal{L} =
	 \sum_{k=i}^{n_*}\log\left(P(S|\mathbf{u}^i)\right) 
+ \sum_{k=i}^{n_*}\log\left(P(\mathbf{u}^i|M)\right)\\
\hskip2cm - n_*\log\left(P(S|M)\right).
\end{multline}
We assume a small core action $J_{\rm core} = 50\kpc\kms$, so $M$
is defined by the parameters ($\alpha_0, \alpha_\mathrm{m}, \beta_0, \beta_\mathrm{m}, J_0,
\FeH_{\mathrm{max}},\sigma,a_r, a_{\phi},b_r, b_{\phi}$).
\subsection[]{Selection function}
There is no selection on $v_\parallel$. A star is selected when its total
proper motion lies within the limits specified in Table \ref{tab:selfunc} and
is excluded  otherwise. Hence the selection function is separable as
\begin{equation}
	\begin{split}
		P(S|\mathbf{u}^i) =& p(S|l,b,s,\mu_l^*,\mu_b,\FeH)\\
	   			      	  =& p(S|l,b)\,p(S|\mu_l^*,\mu_b)\,p(S|s,\FeH).
	\end{split}
\end{equation}

\subsubsection[]{Selection on sky positions}
The selection on $l$ and $b$ depends on the coordinates of the SEGUE-II
plates (which sample the sky sparsely), the completeness fraction per plate,
and whether the Sagittarius stream is excluded. The completeness of a plate
depends strongly on $|b|$ because close to the plane available targets are
numerous so an individual star has a low probability of acquiring a spectrum.

For coordinates within $1.49\degree$ of the centre of a plate that is not
excluded because it intersects the Sagittarius stream, the selection function
equals the completeness fraction for that plate, which as stated earlier, we assume to be independent of apparent magnitude. The completeness fractions
are given by
\cite{xue+15}. In Fig.~\ref{fig:lbsfmaps} the colour scale shows the
selection function in sky coordinates. Near the poles the selection function is large but there are not many stars.
\subsubsection[]{Selection on distance and metallicity}
We use isochrones to tabulate on a grid in $s$ and [Fe/H] the probability
that a star randomly chosen at birth passes the sample's magnitude and colour cuts.
Given our assumption that all halo stars have age $\tau=11\Gyr$, this
involves integrating the IMF over values of the mass $m$ that correspond to the
appropriate absolute
magnitudes
\begin{equation}
p(S|s,\FeH) = \int \diff m \, p(S|m,\tau,s,\FeH).
\end{equation}
We use the $\alpha$-enhanced B.A.S.T.I. isochrones for mass-loss parameter
$\eta=0.4$ \citep{pietrin+06}. We store each isochrone in one-dimensional
interpolants that predict absolute magnitude and colours given the mass.
For given values of $s$ and [Fe/H] we randomly choose 5000 masses from a
uniform distribution that covers the range of masses that correspond to stars
luminous enough to enter the sample, and if a star passes the magnitude and colour cuts,
we add its IMF value to $p(S|s,\FeH)$.

\begin{figure}
\centering
\includegraphics[scale=0.5]{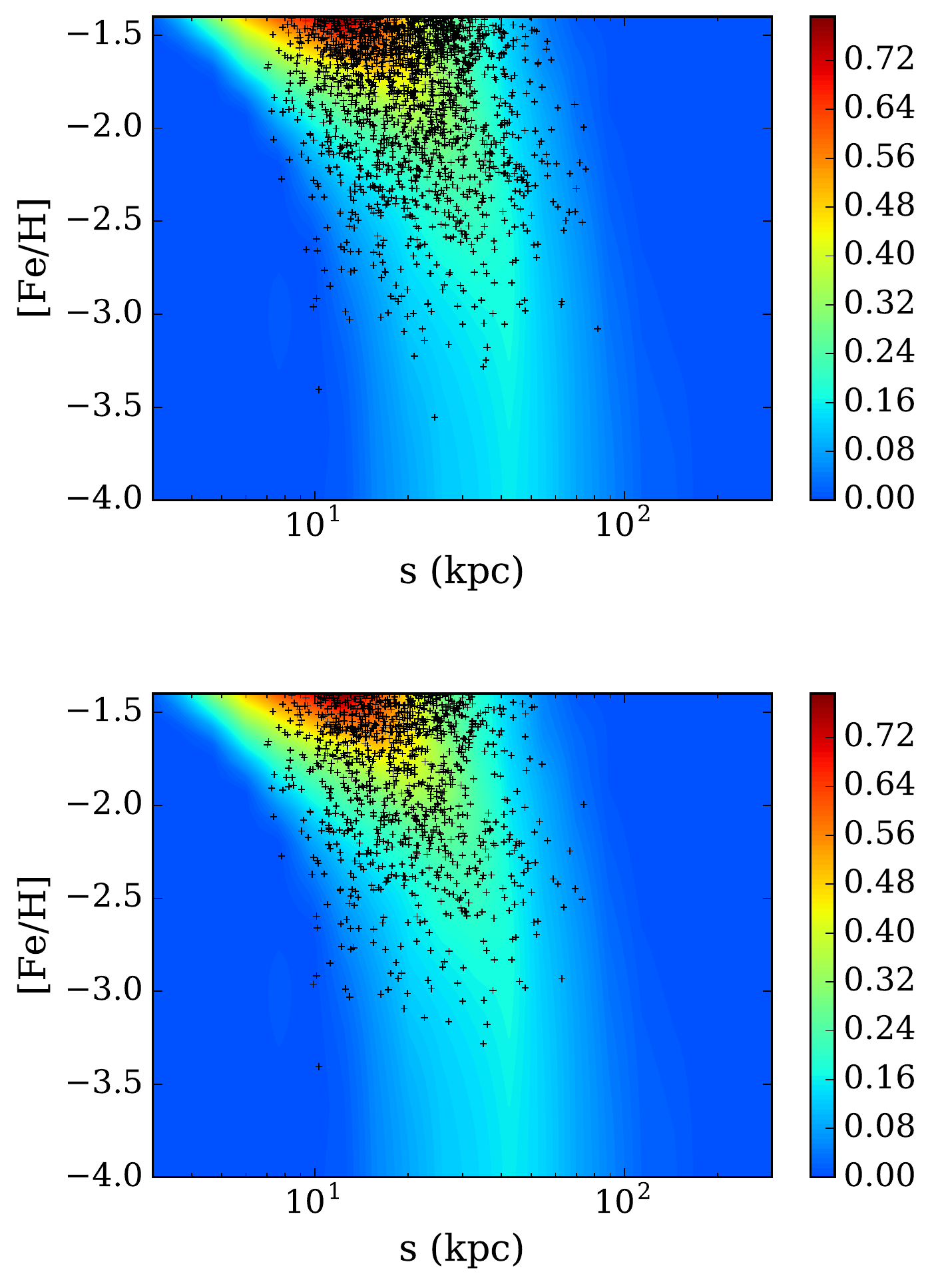}
 \caption{The colour scale shows as a function of distance and metallicity
the selection probability $p(S|s,\FeH)$ with (top) and without (bottom) the
Sagittarius stream included. The symbols mark the locations of observed stars.\label{fig:chemsfmaps}}
\end{figure}
The colour scale in Fig.~\ref{fig:chemsfmaps} shows  $p(S|s,\FeH)$. At any
given value of [Fe/H], $p(S|s,\FeH)$ first rises with distance $s$ beyond the bright apparent magnitude cut-off and then declines sharply when the limiting apparent
magnitude is reached. At any value of $s$,  $p(S|s,\FeH)$ declines when
[Fe/H] is reduced because the colours predicted by the isochrones start falling out of the allowed colour boxes.

The symbols in Fig.~\ref{fig:chemsfmaps} mark the locations of the observed
stars, which are largely, but not entirely, confined to regions of
non-negligible $p(S|s,\FeH)$. The density of stars should, of course, be
proportional to the product of the EDF and the selection function integrated
through a volume.
\subsection[]{Convolution of the EDF with the error distribution}
We neglect errors in sky coordinates and adopt (independent) Gaussian error
distributions for $v_\parallel$, $\mu_l^*$, $\mu_b$, [Fe/H], and $\log s$.
Thus the multi-variate error distribution is
\begin{equation}\label{eqn:convint}
	C^{(7)}(\mathbf{u}^i,\mathbf{u}^{\prime},\mathbf{p}^i) \equiv
	\prod_{l=1}^2\delta(u_l^i-u_l')\prod_{l=3}^{7}C(u_l^i,u_l^{\prime},p_l^i).
\end{equation}
Here true values of observables are indicated by primes and
\begin{equation}
C(u^i,u',p^i)\equiv{\exp\left[{-(u^i-u')^2/(2p^{i2})}\right]\over\sqrt{2\pi} p^i}.
\end{equation}
The
convolution of the EDF with the error distribution of a given star is
\begin{multline}
	P(\mathbf{u}^i|M) = \\
	\int \diff^7\mathbf{u}^{\prime} \, C^7(\mathbf{u}^i,\mathbf{u}^{\prime},\mathbf{p}^i) \, f(\mathbf{\mathbf{x},\mathbf{v},\FeH})) \left|\frac{\partial(\mathbf{x},\mathbf{v},\FeH)}{\partial(\mathbf{u}^{\prime})}\right| .
\end{multline}
The Jacobian determinant here is proportional to $s^4\cos b$
\citep{mcmillan+12}. The
integral is calculated using a fixed Monte Carlo sample of 2000 points to
eliminate the impact of Poisson noise on the likelihood \citep{mcmillan+13}.
The calculation of the integral for all the observed stars is
parallelised using OpenMP over as many cores as are available.
\subsection[]{The normalisation factor}\label{ss:normfactor}
The normalisation of $\mathcal{L}$ is given by
\begin{multline}
P(S|M) = \\
\int\diff^7\mathbf{u}^{\prime}\,P(S|\mathbf{u}^{\prime})\,f(\mathbf{x},\mathbf{v},\FeH)\left|\frac{\partial(\mathbf{x},\mathbf{v},\FeH}{\partial(\mathbf{u}^{\prime})}\right| .
\end{multline}
The selection function could also be convolved with the observed error
distribution as some stars will scatter into the
observational volume, while others scatter out by observational errors. We
do not do this  on account of computational constraints. We approximate the
integrals over sky coordinates by a sum of the values taken by the remaining
five-dimensional integral at the centre of
each included SEGUE plate.  
\begin{figure*}
\centering
\includegraphics[scale=0.6]{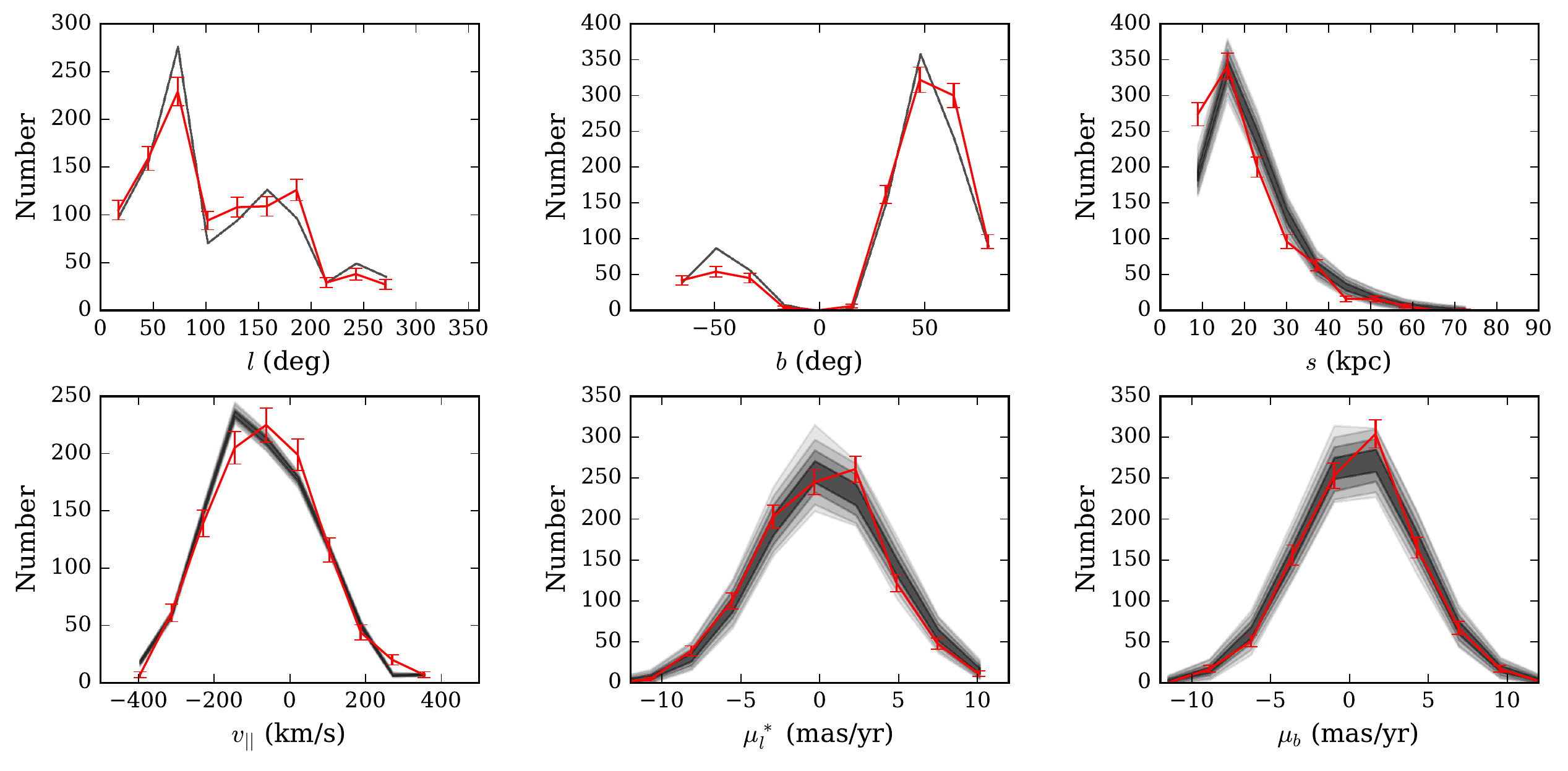}
 \caption{In red: one-dimensional distributions of mock phase-space
observables. The error bars show Poisson errors.
Grey regions: 2000 Monte Carlo resamplings of the data from the error
distributions when the Sagittarius stream is
excluded. Progressively lighter shades of grey indicate $1\sigma$ to $3\sigma$
and 100\% regions for the resampled observables.\label{fig:datafit_1d}}
\end{figure*}

Since $P(S|\mathbf{u}^{\prime})$ is multiplied by  $n_*$ in
equation (\ref{eqn:logL}),  it must be calculated to a high degree
of accuracy \citep{mcmillan+12}. We calculate it to an accuracy of 0.1\%
using the cubature method implemented in the \texttt{Cuhre} routine \citep{hahn05}. The package is highly
versatile, with a range of both Monte Carlo and deterministic algorithms, and
automatic core parallelization.
\subsection[]{Maximising the likelihood}
To derive the probability distribution of the model parameters would require
a Markov Chain Monte Carlo (MCMC) approach. Standard MCMC provides an
estimate of the propagation of random errors and is useful for examining
correlations between parameters. Here systematics
arising because the true EDF is not smooth, our selection function and
potential are imperfect, and the EDF is parameterized, likely dominate the
uncertainties. Hence, we restrict ourselves to finding the EDF that maximises
the likelihood of the data.  We do this using the Nelder-Mead `amoeba'
algorithm \citep{nelder+65}, subject to the following priors. We impose
$0<\alpha(\langle{\FeH}\rangle) \leq \beta(\langle{\FeH}\rangle)$ and
$\beta(\langle{\FeH}\rangle)>3$, where the angled brackets denote a mean.
This ensures a density profile that decreases outwards for all metallicities,
has an outer slope that is either the same as or steeper than the inner
slope, and corresponds to a finite total mass. We require $a_i,b_i>0.2$ to
drive the fitting procedure towards models in which all three actions play a role
in setting the value of the EDF.

\begin{table}
 \centering
  \caption{EDF parameters of various models. Col 1: Nearly isotropic model used to
validate the likelihood methodology. Cols 2 to 4: best-fitting EDF parameters
at various stages when obtaining fits with  the Sagittarius stream excluded.
Col 5: final parameters when  the Sagittarius stream is included. Col 6:
parameters when the overall metallicity distribution comprises the sum of two
Gaussians in $G$.  The * in columns 2 to 6 signifies that the
parameter was fitted during the run. The units of actions are $\kpc\kms$. \label{tab:models}}
\begin{tabular}{lcccccc}
  	\hline
  	Parameter     		   &1		&2		&3		&4		&5    &6\\
  	\hline
  	$J_{\mathrm{core}}$     &50.    	&50.		&50.		&50.		&50.		&50.\\
  	$J_0$				   &5000.	&4979.*	&5000.	&5000.  &5000.	&5000.\\ 
  	$\alpha_0$      		   &3.50		&1.39*	&2.27*	&2.34*	&1.81*  &2.34\\
  	$\alpha_\mathrm{m}$		       &-0.50	&0.0	&-0.35*	&-0.44* &-0.26*	&-0.44\\
  	$\beta_0$		       &5.00		&4.51*	&5.24*	&5.97*	&5.96*  &5.97\\
  	$\beta_\mathrm{m}$			   &-0.50	&0.0	&-0.29*	&-0.42*	&-0.41*	&-0.42\\
  	$a_r$				   &1.50		&1.5	&1.50	&0.24*	&0.22*  &0.24\\
  	$a_{\phi}$			   &0.75		&0.45	&0.45	&0.87*	&0.90*  &0.87\\
  	$a_z$				   &0.75		&1.05	&1.05	&1.89*	&1.88*  &1.89\\
  	$b_r$				   &1.20		&0.6	&0.6	&1.36*	&1.14*  &1.36\\
  	$b_{\phi}$			   &0.90		&0.80	&0.80	&0.73*  &0.81*  &0.73\\
  	$b_z$				   &0.90		&1.6	&1.6	&0.91*  &1.05*  &0.91\\
  	$w	$			   &1.00		&1.0	&1.0	&1.0	&1.00   &0.49*\\
  	$\FeH_{\mathrm{max}}$   &-0.80	&-0.80	&-0.78*	&-0.77*	&-0.76* &-0.77\\
  	$\sigma$				   &0.35		&-0.35	&-0.39*	&0.40*	&0.40*	&0.57*\\
  	$w'$					   &0.00	    &0.00	&0.00	&0.00	&0.00	&0.51*\\
  	$\FeH_{\mathrm{max}}'$  &-	    	&-		&-		&-		&-		&-1.42*\\
  	$\sigma'$			   &-		&-		&-		&-		&-		&0.36*\\

  	\hline
  \end{tabular}
\end{table}

As stated earlier, we fix $J_{\mathrm{core}}$ to $50\kpc\kms$. On account of
the large number of parameters to fit, we seek the maximum-likelihood model
in three stages. We start by fitting the base power-law indices ($\alpha_0$
and $\beta_0$) and the scale action ($J_0$) with the other parameters set to
sensible values. Specifically, we set the metallicity indices ($\alpha_\mathrm{m}$ and
$\beta_\mathrm{m}$) to zero and the lognormal parameters ($\mathrm{[Fe/H]_{\mathrm{max}}}$ and
$\sigma$) to the values obtained by fitting the global metallicity
distribution. We set the action weights ($a_r$, $a_{\phi}$, $b_r$, and
$b_{\phi}$) to the values given in Col.~1 of Table~\ref{tab:models}. These
parameters produce a
flattened density ellipsoid with an orbital structure that becomes more
radially anisotropic outwards.  The initial amoeba for this first fit is kept
relatively large to allow a thorough scan of the three-dimensional parameter
space. In the second step, $J_0$ is fixed to the value found in the first
step and the other parameters are varied with the
exception of the action weights, $a_r$, $a_{\phi}$, $b_r$, and $b_{\phi}$.
For this step the initial amoeba is narrower in $\alpha_0$ and $\beta_0$ than
in the other parameters.  In the final step, we allow the action weights to
vary, again with the initial amoeba narrower in parameters already fitted and
wide in the action weights. The outputs of each stage are given in
Table~\ref{tab:models}.
\begin{figure}
\centering
\includegraphics[scale=0.6]{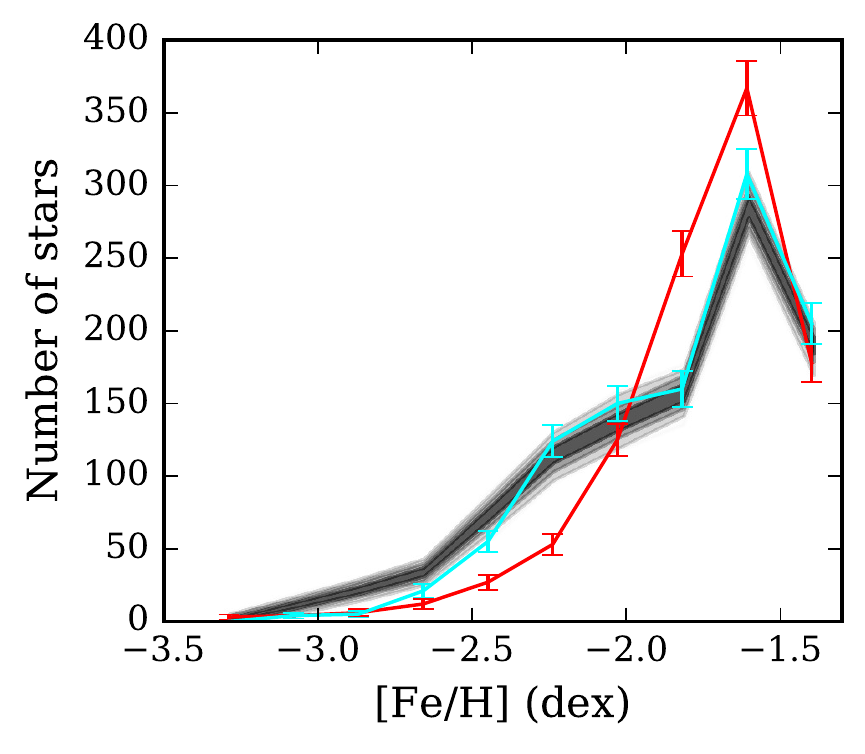}
 \caption{In red and grey: as Fig.~\ref{fig:datafit_1d} but
 for metallicities when  the EDF is defined in terms of a single Gaussian in
$G$. In cyan: the distribution of mock metallicities when the
EDF is defined by two Gaussians in $G$.\label{fig:fehfit_1d}}
\end{figure}

To validate the likelihood methodology, we used a sampling-rejection
algorithm to draw 1000 actions from the EDF with parameters representing an
approximately isotropic, round model (Col.~1 of Table ~\ref{tab:models}). We
complement these actions with angles uniformly distributed on $(0,2\pi)$ and
used torus mapping \citep{binney_mcmillan15} to convert $(\vtheta,\vJ)$ to
the first six coordinates in equation ~\eqref{eqn:obs}. We then added 5\%
Gaussian errors to these mock observations, and used them to fit an EDF
using the likelihood methodology described above. We repeated this procedure
ten times to study how the fitted parameters scattered around the true
values. The true values were captured in the range of best-fit parameters
derived from the mock observations.

\begin{figure*}
\centering
\includegraphics[scale=0.7]{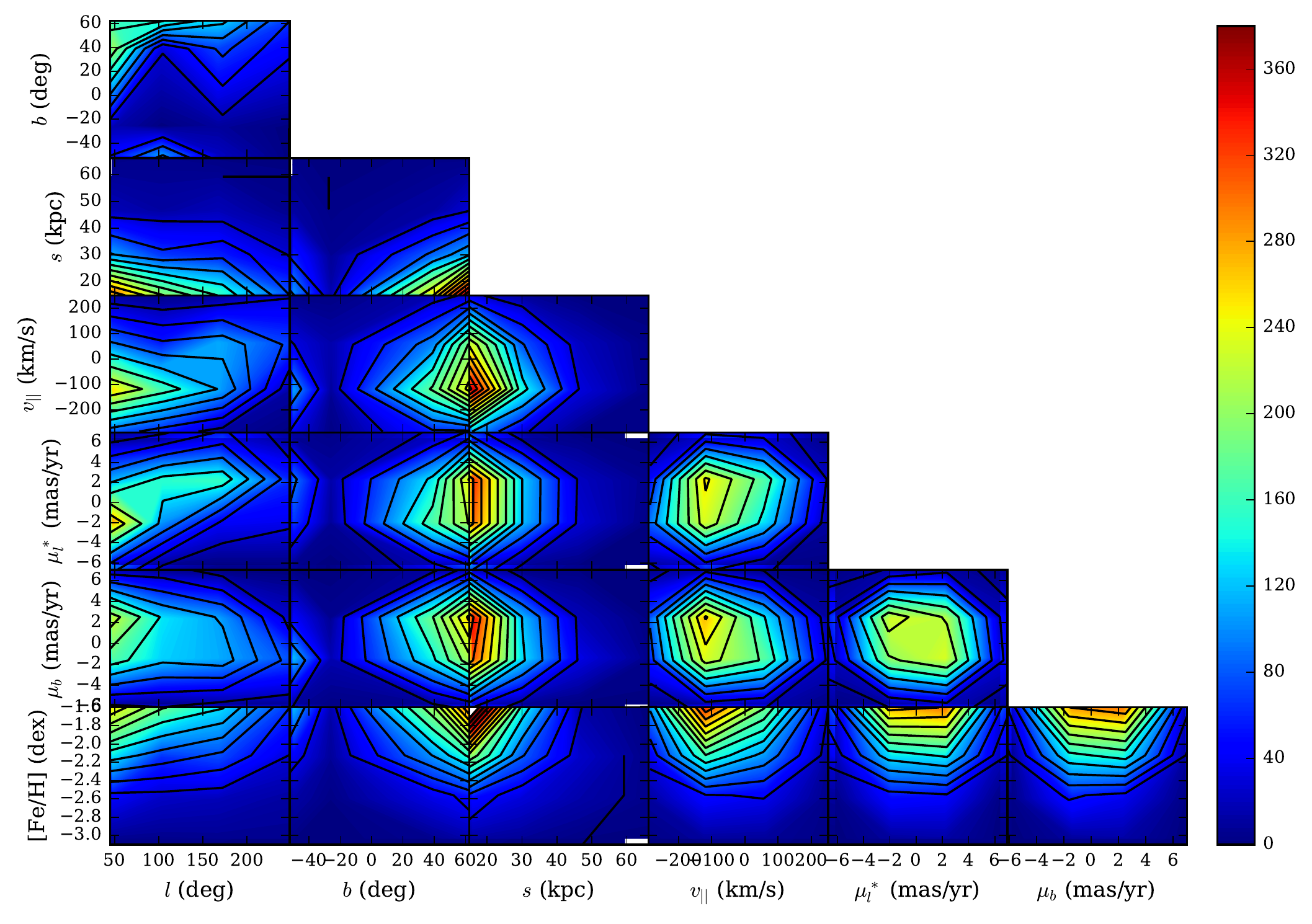}
 \caption{Colour-filled contours: two-dimensional
distributions of mock observables when the Sagittarius
stream is excluded. Black contours: distributions of the measured
observables.
\label{fig:datafit_2d}}
\end{figure*}
\section[]{Results}
Here we discuss the parameters of the best-fitting EDF, the quality of the
fit, and the structure of this EDF in action and phase space.
\subsection[]{The best-fit parameters} 
In Table \ref{tab:models} we give the best-fit parameters without
(Cols.~2--4, ~6) and with (Col.~5)
the Sagittarius stream.  The parameters are
similar in the two cases.  The halo's inner region has velocity ellipsoids
that are strongly elongated in the radial direction ($a_r\ll a_\phi,a_z$) and
such that the azimuthal velocity dispersion is larger than the dispersion
perpendicular to the Galactic plane ($a_\phi< a_z$). The inner halo is flattened as a
consequence. The outer halo is a nearly spherical system with almost
isotropic velocity ellipsoids ($b_r\simeq b_\phi\simeq b_z$). The halo's
density profile always steepens around the radius corresponding to $J_0$
($\alpha_0\ll\beta_0$). The principal effects of removing the stream are (i)
to strengthen the dependence upon [Fe/H] of the inner density profile
($\alpha_\mathrm{m}=-0.44$ versus $-0.26$), and (ii) to reduce the radial anisotropy
of the outer region ($b_r=1.36$ versus 1.14).

\subsection[]{Fits to the observables}
We assess the fit of the model to the observables when the stream is excluded
by generating a mock catalogue, using a sampling-rejection method and taking
into account the selection function and errors. The sampling-rejection method
involves constructing a proposal density that is a function of the
coordinates to be sampled. Our proposal density is the product of Gaussian
distributions in each of $x$, $y$, $z$, $v_x$, $v_y$, $v_z$, and $G$. We pick
a point in the seven-dimensional space by sampling the proposal density, and
evaluate the EDF there. We compute the point's heliocentric coordinates
$(l,b,\log s,\mu_l^*,\mu_b,v_\parallel)$ and Gaussian noise is added to them
and to $G$. The selection function is evaluated at the point just derived and
multiplied by the previously derived value of the EDF. If the ratio of this
product to the proposal density is greater than another uniformly generated
random number, the mock observables are added to the sample.
\begin{figure}
\includegraphics[scale=0.5]{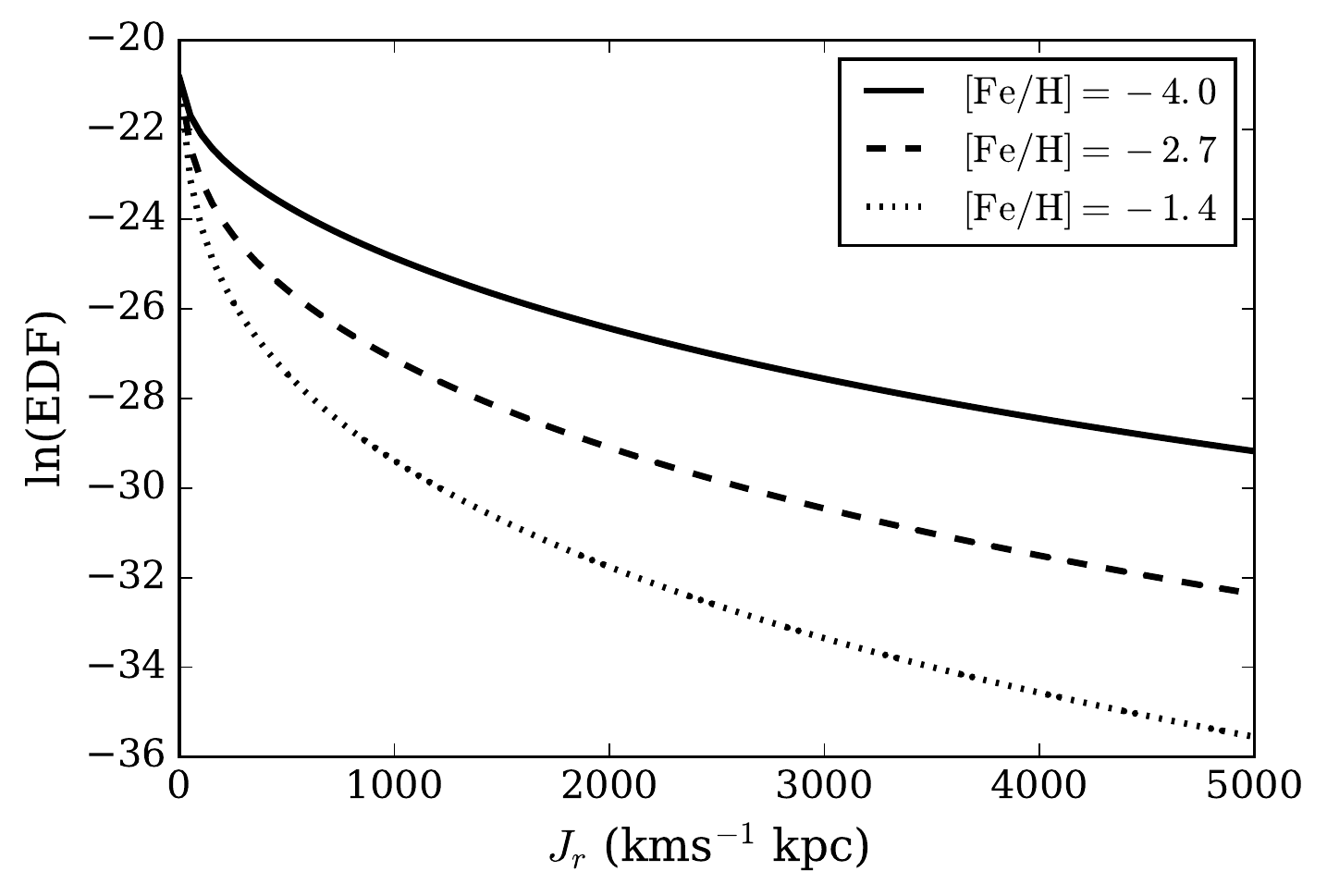}
	\caption{The value of  the best-fitting EDF
 along the line $(J_r,1.5J_r,0.8J_r)J_r$ through action space for three metallicities. \label{fig:model_edf1d}}
\end{figure}

The red points joined by lines in Figs.~\ref{fig:datafit_1d}
and \ref{fig:fehfit_1d} show histograms of the mock catalogue. The region
covered by analogous histograms of 2000 resamplings of the error
distributions of the measured
observables are shown by the grey regions. In plots for observables with
small errors, such as $l$ and $b$ and $v_\parallel$, the grey regions form
fairly well defined curves. In plots for observables with large errors,
namely $(s,\mu_l^*,\mu_b,G)$, the grey regions fill out a region of
significant width. In every case but one, the red curve lies near the centre
of this swathe, indicating that the EDF and selection function are together
doing a good job in the sense that the data would be reasonably likely to
occur if the EDF and selection function accurately evaluated the
probabilities. The exception to this statement is the metallicity. Although
the general shape of the observed metallicity DF is captured, the fit is
poor. So we generate a new mock catalogue of $\FeH$ by adding a second
lognormal component to $f_\mathrm{m}$
\begin{equation}
f_{\mathrm{m}}(G,G') =
	w\e^{G}\frac{\e^{-\frac{G^2}{2\sigma^2}}}{\sigma\sqrt{2\pi}} 
+ w'\e^{G'}\frac{\e^{-\frac{G'^2}{2\sigma'^2}}}{\sigma'\sqrt{2\pi}},
\end{equation}
where $G$ and $\sigma$ are as before, $w+w' = 1$, and $G'$ is associated with
a new $\FeH_{\textrm{max}'}$ and $\sigma'$. A non-rigorous exploration of the
space finds a much improved fit to the observed metallicities for parameters
given in Col.~6 of Table~\ref{tab:models} and illustrated in
Fig.~\ref{fig:fehfit_1d}. These parameters generate peaks at
$\FeH\sim-1.5$ and $\FeH\sim-2.3$. The exercise of fitting the revised EDF
with a second Gaussian in $G$ ab initio is left for a future paper, in which
substructures are sought in chemodynamical space.

Fig.~\ref{fig:datafit_2d} compares histograms for the joint distribution of
pairs of phase-space observables $(b,l)$, $(s,l)$, $(\FeH,l)$, etc.,
when the Sagittarius stream is excluded and the EDF is
defined in terms of a single Gaussian in $G$. The colour-filled contours show
distributions of mock observables, while the black contours show the
distributions of measured observables. In general there is good agreement
between the mock and observational distributions.

\subsection[]{The distribution of stars in action space and metallicity}
\begin{figure}
\centering
\includegraphics[scale=0.4]{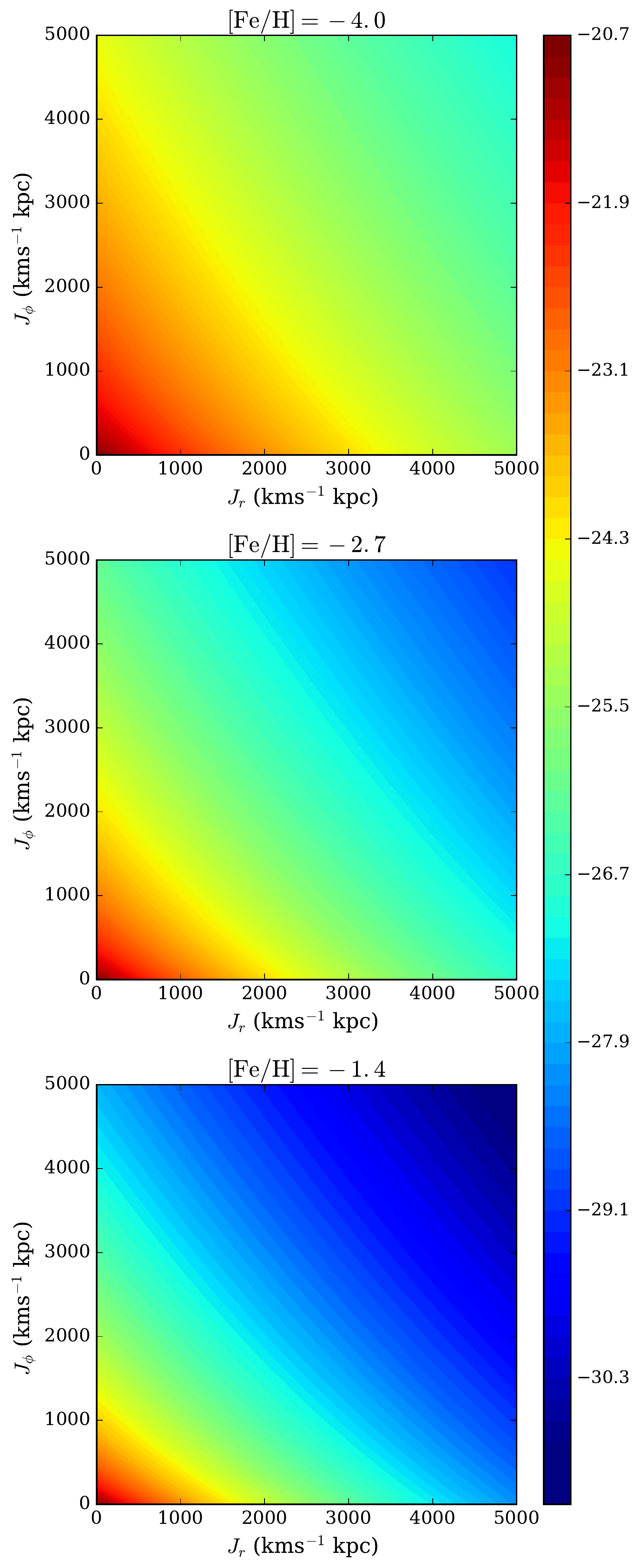}
	\caption{Best-fitting EDF as a function of radial action $J_r$ and angular
momentum $J_{\phi}$, with fixed $J_z=100\kpc\kms$. The colour scale shows logarithmic density.
\label{fig:model_edf2d}}
\end{figure}
\begin{figure}
\centering
\includegraphics[scale=0.7]{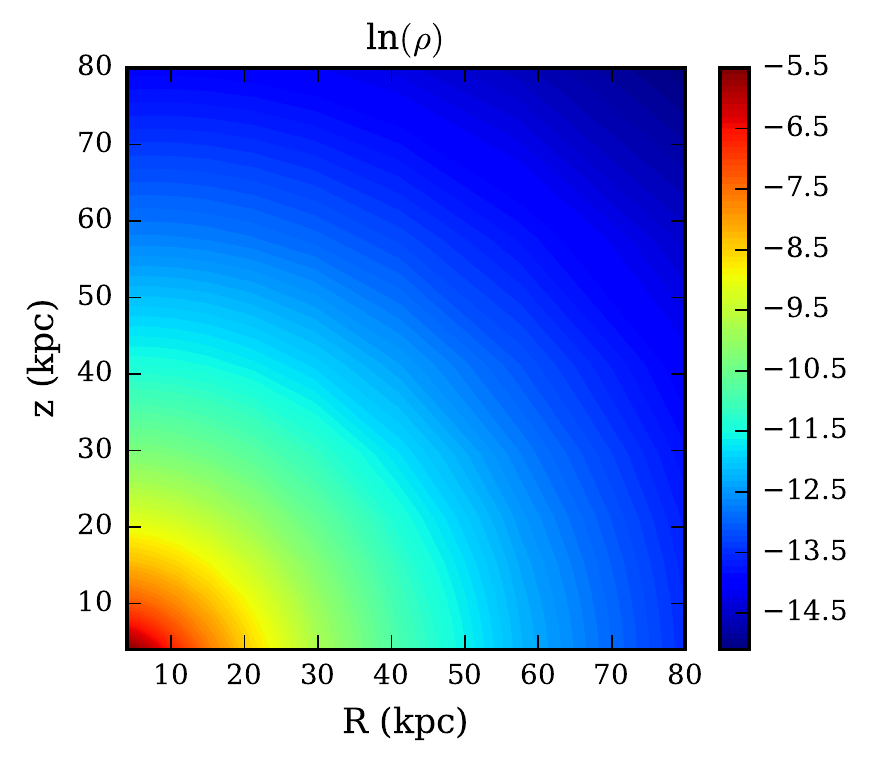}
	\caption{Logarithmic density moment of the best-fitting EDF. \label{fig:model_densitymoment}}
\end{figure}
\begin{figure}
\centering
\includegraphics[scale=0.65]{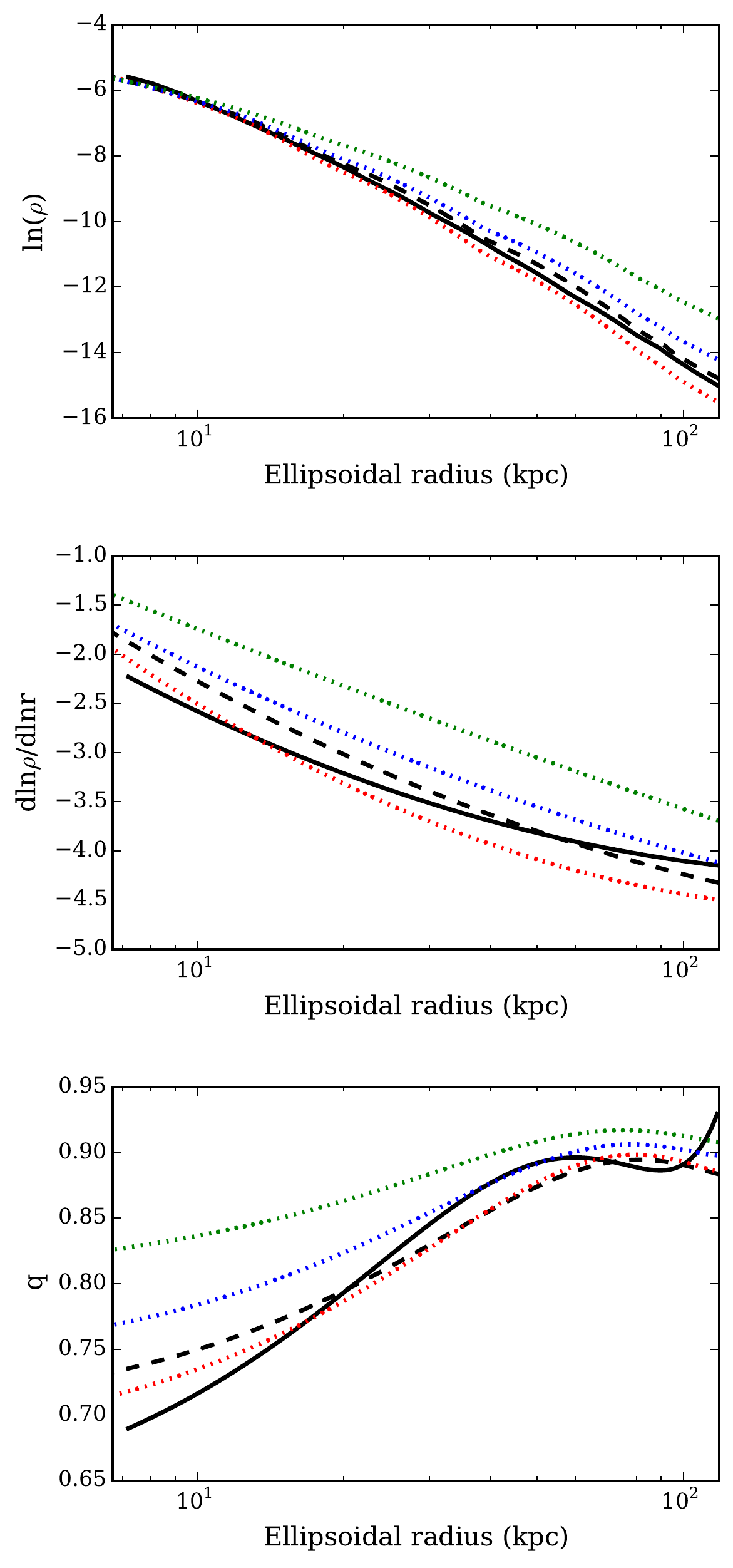}
	\caption{One-dimensional profiles of logarithmic density, logarithmic density
gradient, and axis ratio of the density ellipsoids when: (i) the
Sagittarius stream is excluded (black, solid); (ii) the Sagittarius stream is
included (black, dashed); (iii) the Sagittarius stream is excluded and $\FeH =
-1.4$ (red, dotted); (iv) the Sagittarius stream is excluded and $\FeH = -2.7$
(blue, dotted); (iii) the Sagittarius stream is excluded and $\FeH = -4.0$
(green, dotted). \label{fig:model_density1d}}
\end{figure}
Fig.~\ref{fig:model_edf1d} shows for metallicities $\FeH=-4.0$, $-2.7$, and
$-1.4$ plots of the best-fitting $f(\vJ,\FeH)$ along the line in action space
given by $J=(J_r,\,1.5J_r,\,0.8J_r)$, thus probing similar halo orbits at
different energies. The values of $f$ have been scaled to agree at $J_r=0$.
We see that $f$ declines more steeply with $J_r$ for metal-richer stars,
implying that these stars are tightly confined near the origin of action
space.

Each panel of Fig. \ref{fig:model_edf2d} shows for a different metallicity
the value of $f(\vJ)$ in the action-space plane $J_z=100\kpc\kms$, again in the case the stream is excluded. Contours of
constant $f(\vJ)$ are approximately straight lines because they are similar
to the lines on which either $h(\vJ)$ or $g(\vJ)$ are constant.  At small
values of $|\vJ|$, the slopes of the contours are $\simeq -a_r/a_\phi$, and at
large $|\vJ|$ they tend to $-b_r/b_\phi$. Comparing the top panel for
$\FeH=-4$ with the bottom panel for $\FeH=-1.4$, the closer confinement of
metal-rich stars to the origin of action space is evident.
\subsection[]{Moments of the best-fitting DF}

Fig.~\ref{fig:model_densitymoment} shows the density $\rho(R,z)$ predicted by
the best-fitting EDF when the Sagittarius stream is excluded. The flattening
of the halo can be discerned. By fitting ellipses to the isodensity
curves in this figure we obtain the radial density profiles shown in the top
panel of Fig.~\ref{fig:model_density1d}. The full black line shows the
profile obtained by excluding the Sagittarius stream, while the dashed black
line shows the profile obtained by including the stream. The difference
between these two profiles is barely discernible. The red dotted line shows
the radial profiles of the most metal-rich stars ($\FeH=-1.4$), while the
green dotted curve shows that of the most metal-poor stars ($\FeH=-4$). Again
the increase in the compactness of the halo as [Fe/H] is increased is evident.
The bottom panel of Fig.~\ref{fig:model_density1d} shows the axis ratios of
the entire halo (black curves) and its mono-abundance components (coloured
dotted curves). The inner halo is more flattened ($q\simeq0.69$) than the
outer halo ($q\simeq0.93$), and flattening decreases as one moves from the
most metal-rich to the most metal-poor component.  The central panel of
Fig.~\ref{fig:model_density1d} shows plots of logarithmic density gradients.
The gradient of the overall profile (full curve) steepens from $\sim-2.3$ in
the inner halo to $\sim-4.1$ in the outer halo.

We characterise the mixed velocity moments by calculating the tilt in the
velocity ellipsoid using \citep{smith+09}
\begin{equation}
	\tan(2\alpha_{ij}) = \frac{2\sigma_{ij}}{\sigma_{ii} - \sigma_{jj}},
\end{equation}
where $i,j=r,\phi,\theta$ and $\alpha_{ij}$ is the angle between the major
axis of the projection of the velocity ellipsoid in the $(i,j)$ coordinate
plane and a coordinate direction. We find the tilt to be close to zero for
all projections, which means the velocity ellispoids all point towards the centre.

\begin{figure*}
\centering
\includegraphics[scale=0.8]{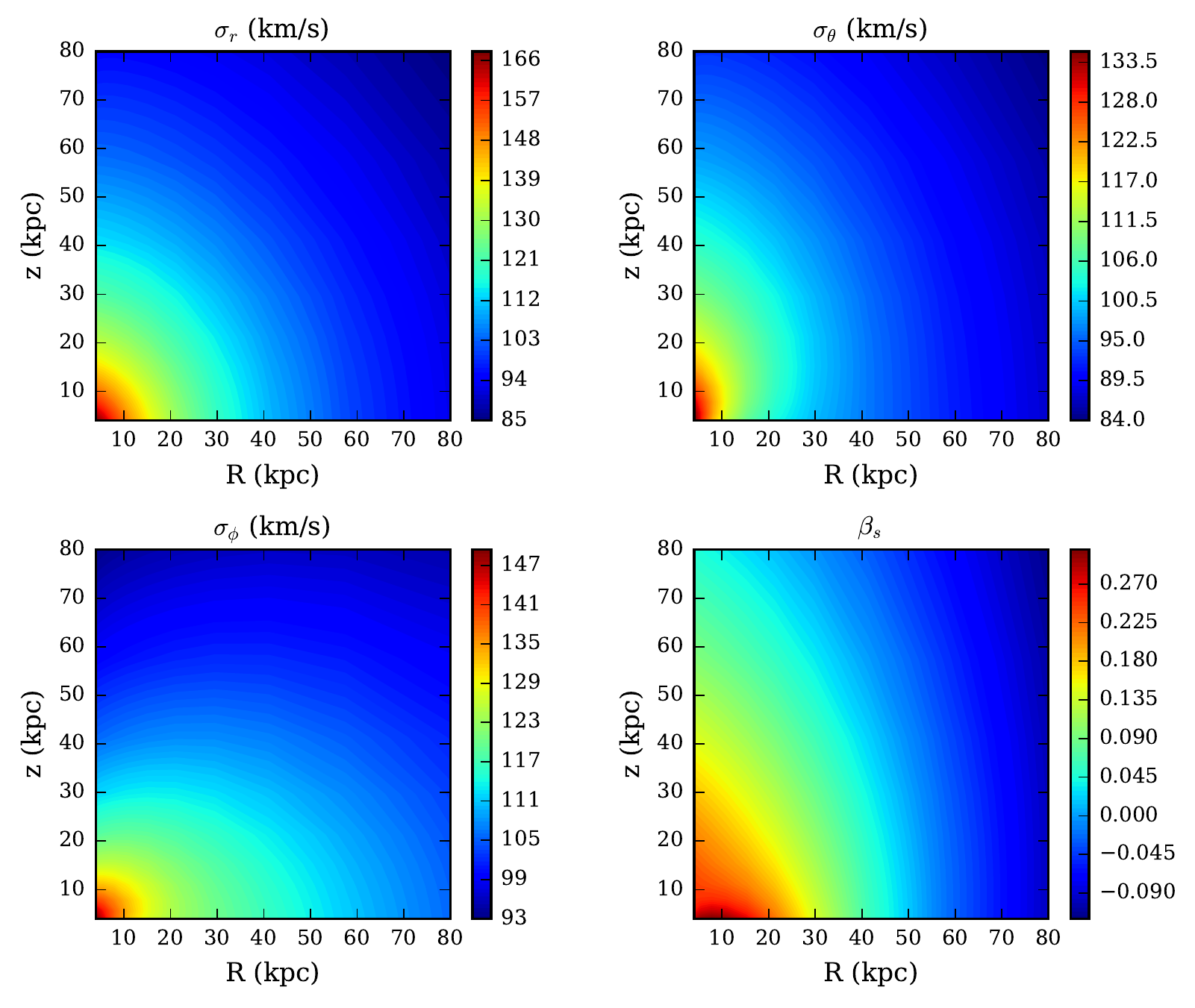}
 \caption{Velocity dispersions predicted by the
 best-fitting EDF when the Sagittarius
stream is excluded. Going from the top left clockwise: spherical radial
velocity dispersion, spherical angular velocity dispersion, the spherical anisotropy parameter, and spherical azimuthal velocity dispersion.
\label{fig:model_dispmoments}}
\end{figure*}

\begin{figure}
\centering
\includegraphics[scale=0.55]{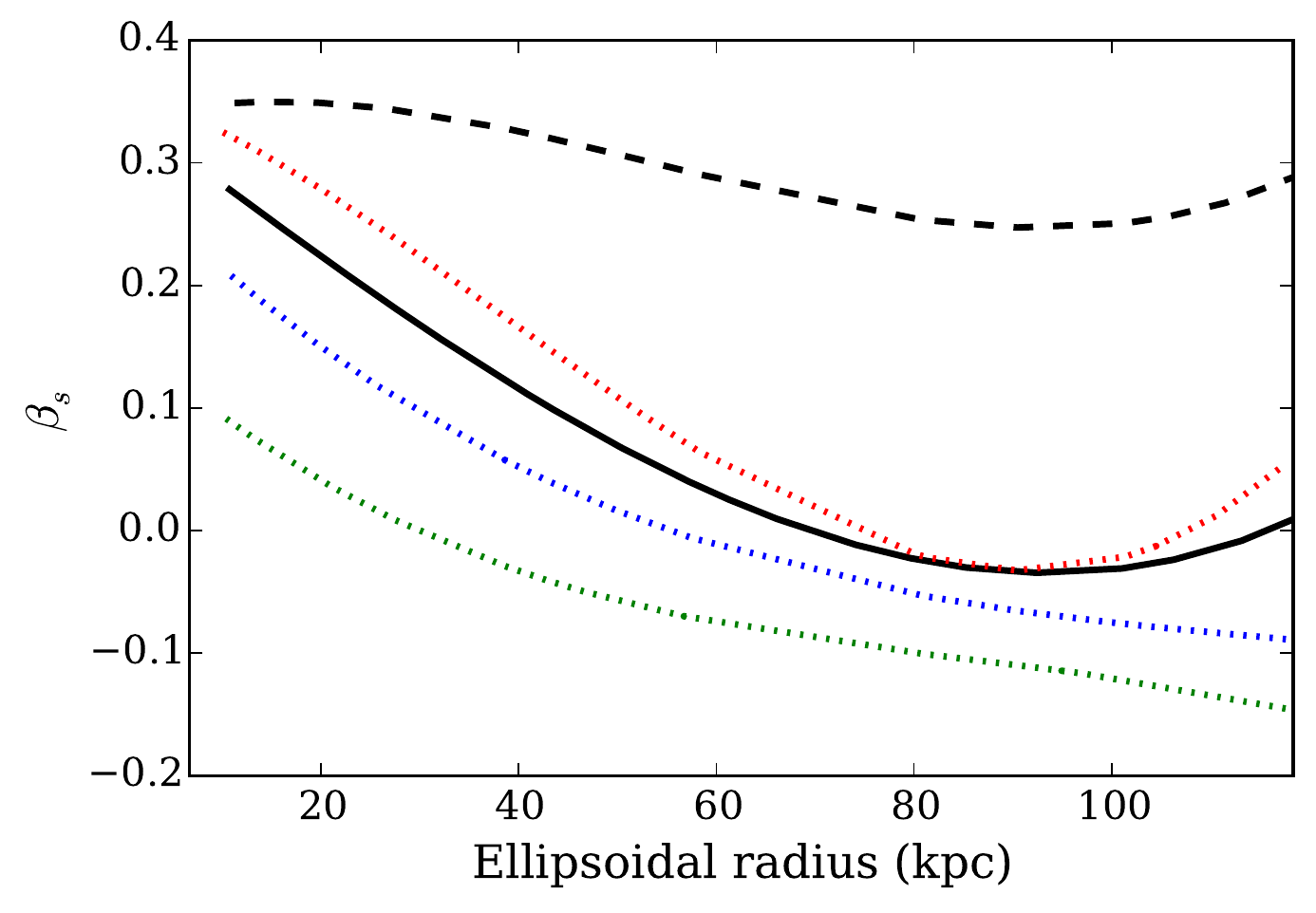}
	\caption{Spherical anisotropy parameter as a function
of ellipsoidal radius when:  (i) the Sagittarius stream is excluded
(black, solid); (ii) the Sagittarius stream is included (black, dashed); (iii)
the Sagittarius stream is excluded and $\FeH = -1.4$ (red, dotted); (iv) the
Sagittarius stream is excluded and $\FeH = -2.7$ (blue, dotted); (iii) the
Sagittarius stream is excluded and $\FeH = -4.0$ (green, dotted).}
\label{fig:model_beta1d}
\end{figure}

Fig.~\ref{fig:model_dispmoments} shows that contours of constant $\sigma_r$
are mildly elongated, and those of constant $\sigma_\theta$ are strongly elongated along
the $z$ axis. By contrast, contours of constant $\sigma_\phi$ are flattened. 
The fourth panel
of Fig.~\ref{fig:model_dispmoments} plots the spherical velocity-dispersion
anisotropy parameter
\begin{equation}
\beta_{\rm s}= 1 - \frac{\sigma_{\theta}^2 + \sigma_{\phi}^2}{2\sigma_r^2}.
\end{equation}
This is generally positive, implying radial bias, with the highest values
$\beta_{\rm s}\simeq0.3$ attained at $R\la12\kpc$ on the equatorial plane.
The anisotropy varies between moderate radial bias in the inner halo
($\beta_\mathrm{s}\sim0.27$) to near isotropy ($\beta_\mathrm{s}\sim-0.09$)
in the outer halo. To compare anisotropy profiles when the Sagittarius stream
is excluded and included, we calculated a one-dimensional spherical
anisotropy profile by averaging over bins in ellipsoidal radii.
Fig.~\ref{fig:model_beta1d} shows the resulting profiles of $\beta_{\rm s}$ for
the whole halo (black lines) and for the halo binned by metallicity (dotted
coloured lines). Anisotropy tends to decrease outwards. Including the
Sagittarius stream (broken black line) changes $\beta_{\rm s}$ to roughly
constant between 0.28 and 0.35 throughout. Since the metal-rich stars are
concentrated at low actions where the anisotropy is largest, at any given
radius anisotropy increases with metallicity, being everywhere below
$\beta_{\rm s}\simeq0.1$ at $\FeH=-4$.

\begin{figure}
\centering
\includegraphics[scale=0.8]{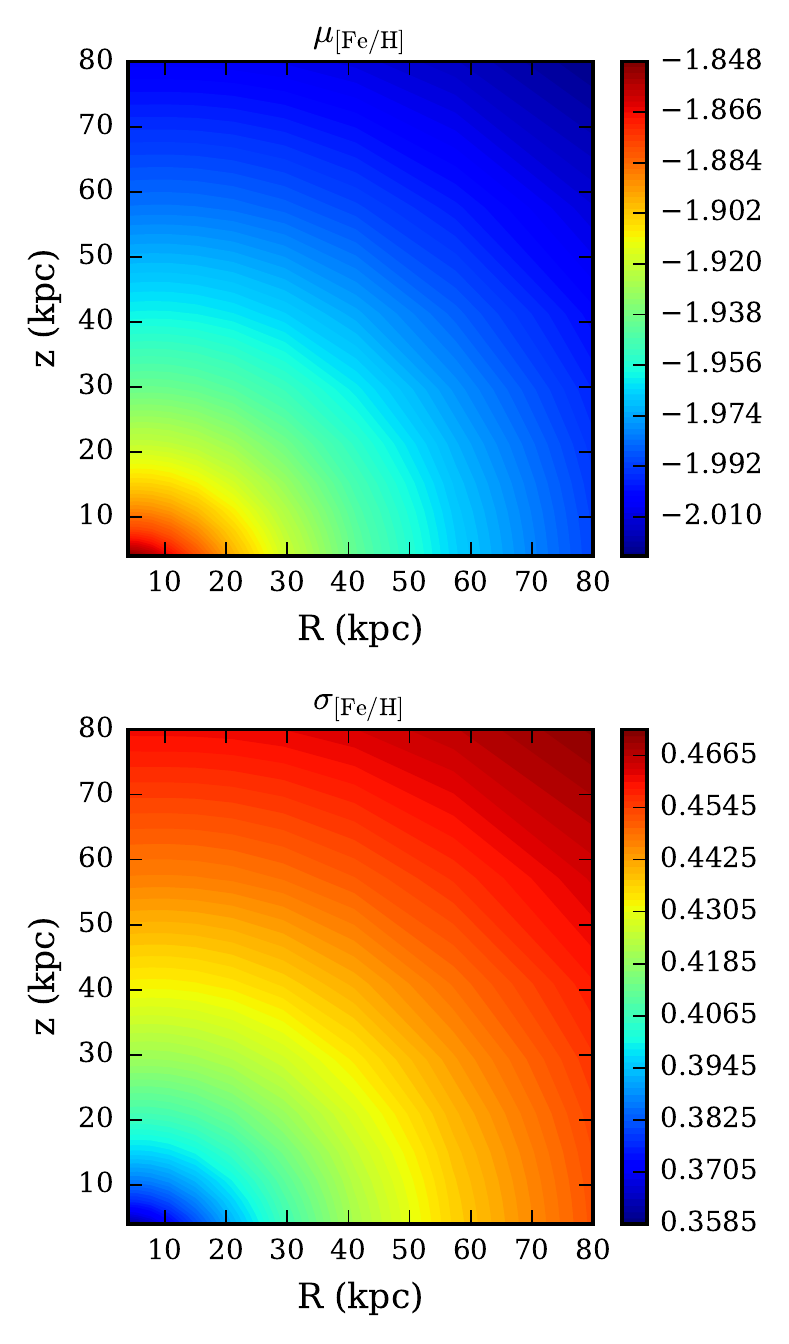}
 \caption{Metallicity moments of the best-fitting EDF when the Sagittarius
stream is excluded. Top: mean metallicity. Bottom:  the
dispersion in metallicities. \label{fig:model_fehmoments}}
\end{figure}

The top panel of Fig.~\ref{fig:model_fehmoments} shows the mean metallicity
as a function of $(R,z)$. Surfaces of constant $\langle\FeH\rangle$ are
moderately flattened, and $\langle\FeH\rangle$ declines outwards from
$\sim-1.85$ near the centre to $\sim-2.01$ at the largest radii, the typical
metallicity gradient being $\sim-0.0014\,$dex/kpc. The lower panel of
Fig.~\ref{fig:model_fehmoments} shows the dispersion in $\FeH$, which
increases outwards from $\sim0.36$ to $\sim0.47$. Hence at any radius the
range of metallicities represented spans the range covered by the mean
metallicity over a full $80\kpc$.

\section[]{Discussion}
\subsection[]{Comparison with previous work}
\subsubsection[]{Radial density profile}
% Change in number of stars with radius
The density of stars decreases with radius with a power-law slope of $\sim
-2$ in the inner halo, steepening to $\sim-4$ in the outer halo. The halo's
axis ratio increases from $\sim0.7$ in the inner halo to $\sim0.9$ in the
outer halo. The findings do not depend on whether we include or omit the
Sagittarius stream, and are consistent with the profile previously determined
for K giants \citep{xue+15}. BHBs also have a density profile that becomes
steeper with radius but the logarithmic slope is claimed to vary much more
rapidly \citep{deason+11}, with potentially a third steeper power-law index
required to describe the density profile beyond $50\kpc$ \citep{deason+14}.
We cannot comment on the sharpness of the transition from the inner to the
outer slope in the K giants because this is fixed by our adopted form of EDF.
By adding an additional parameter to the EDF, it would be straightforward to
vary the sharpness of the transition analogously to how \cite{zhao+96}
generalised the \cite{nfw+96} profile, but we have not explored this avenue.

\cite{carollo+07} inferred the global structure of the halo
from the kinematics of stars now located within $\sim4\kpc$ of the Sun. They
argued that the sample divided into ``inner'' and ``outer'' populations. The
MDF of the inner population peaks at $\FeH\sim-1.6$, while that of the outer
population peaks at $\FeH\sim-2.2$. From local kinematics of a larger sample
\cite{carollo+10} inferred that the inner population has a steeper, more
flattened density profile ($\rho\sim r^{-3.2}$, $q\sim0.6$) than the outer
population ($\rho\sim r^{-1.8}$, $q\sim1$).  Our EDF is consistent with these
findings as regards the peak metallicities and flattenings of the two
populations.  However, both the samples and the analysis method of
\cite{carollo+07} and \cite{carollo+10} differ sharply from ours. Their
dynamics was restricted to a spherical potential, and  their
samples were local and contained a mix of stellar types.
\cite{schonrich+14} argue that their evidence for dichotomy in the halo is an
artefact arising from their failure to model their sample's
selection function.
\subsubsection[]{The velocity ellipsoid}
% Tilt of velocity ellipsoid
We find the velocity ellipsoids to be aligned with spherical
polar coordinates, which may be the consequence of assuming a spherical dark matter halo. Previous studies have found spherical alignment
\citep[e.g.][]{gould+03,smith+09}, but not in the context of a fully
dynamical model. 

%
% Anisotropy
The velocity ellipsoid is radially biased in the inner halo, but when the
Sagittarius stream is excluded, we find it to be nearly isotropic in the
outer halo. Studies of BHBs have generally found the velocity ellipsoid to be
more radially biased \citep{deason+12a,williams+15b}, although from the same
sample of BHBs \cite{hattori+13} reached a similar conclusion to our finding
for the K giants. \cite{carollo+10} also find the outer-halo
stars to be on less eccentric orbits. 

Simulations predict a higher degree of radial anisotropy for stars that have
been accreted by the halo through minor mergers \citep{bullock+05,abadi+06}.
The simulations may be unreliable, but it is also true that the anisotropy in
the outer halo is more uncertain than in the inner halo because stars in the
outer halo are less well represented in the data and their tangential
components of velocities are very uncertain.

\subsubsection[]{Metallicity components}
Two lognormal components are required to capture adequately the metallicity
DF, even when the dependence of the selection function on metallicity is
taken into account. An ad-hoc exploration of the parameters of a DF defined
by two Gaussians finds a metal-richer component peaking at $\FeH\sim-1.5$, with
an upper limit at $\FeH\sim-0.8$, and a metal-poorer component peaking at
$\FeH\sim-2.3$, with an upper limit at $\FeH=-1.4$. \cite{carollo+07},
\cite{chen+14}, \cite{an+15}, and \cite{xue+15} reached similar conclusions, based on very different samples and technqiues. Of these, only \cite{an+15} and \cite{xue+15} consider selection effects.

\subsubsection[]{Dependence of phase-space structure on metallicity}
The metal-rich stars are more tightly confined to the region around the
origin of action space than the metal-poor stars. As a consequence, the
radial density profile of the metal-rich stars is steeper than that of the
metal-poor stars, leading to a weak metallicity gradient
of $\sim-0.0014\,$dex/kpc. This result is consistent with the
analysis of K giants by \cite{xue+15}, but takes the work
further by demonstrating dynamical consistency between the density and
dispersion profiles in a realistic Galactic potential. The literature regarding
determination of gradients in other stellar populations
\citep[e.g.][]{carollo+07,peng+12,chen+14,allende+14,schonrich+14,an+15} shows no 
consensus on the existence of a gradient. The lack of consensus may arise
from improper treatment of the selection function in some cases, with
metal-poorer stars being preferentially detected at larger distances
\citep{schonrich+14}. In other cases, it may reflect 
genuine differences between stellar populations, again emphasising the need to
employ EDFs, and to extend them to cover other characteristics of stars. 

A gradient arises naturally from metallicity gradients in the satellites
digested by the Galaxy: the outer, more metal-poor stars are stripped at
large radii, and the more metal-rich core at smaller radii.  Direct evidence
for this effect is provided by the metallicity gradient along the Sagittarius
stream \citep{chou+07}.  Moreover, dynamical friction drags the most massive
and metal-rich satellites inwards, while the less massive ones experience
less friction and are tidally shredded at large radii.

The dispersion in [Fe/H] increases with radius, indicating that a wider range
of metallicities is contributing to the outer halo. Since metal-rich stars
can be found only in more massive satellites, this finding suggests the outer
halo is produced by stripping a more heterogeneous group of satellites than
produced the inner halo. However, we cannot discount the possibility that
some of the metal-richer stars now found at large radii have been scattered
onto energetic orbits by the Galaxy's bar.

At all metallicities, the radial bias of the velocity ellipsoids decreases
outwards.  Since the metal-richer stars reside deeper in action space, they
have a more radially biased velocity ellipsoid than the metal-poorer stars.
This result mirrors conclusions drawn by \cite{carollo+10}
from a sample of mixed stellar types, and by \cite{hattori+13} from BHBs.

\subsection[]{Uncertainties in the analysis}
\subsubsection[]{Impact of substructure}
Removing the Sagittarius stream diminishes the radial bias of the velocities
in the halo. We do not know how substructures that we have not masked impact
our assessment of the halo's structure. However, the ability to reproduce
adequately the phase-space observations after excluding the stream implies
that the current data for K giants is too sparse in phase space to resolve
the halo's substructure. Indeed, the latter has only been detected in counts
of dwarf stars, which are much more numerous than K giants. It seems that our
EDF gives a satisfactory account of the structure of the halo on the medium
to large scales that can currently be probed with K giants.
\subsubsection[]{Stellar population assumptions}
Our evaluation of the metallicity-distance selection function depended on the
assumption of a single (old) age, and on relations from isochrones between
the age, mass, and metallicity of a star and its luminosity in various
wavebands. Systematic errors arising from faulty isochrones are difficult to
assess.  However, the assumption of a single age for the K giants is likely
to have the following impact. Isochrones for older stars result in very
similar selection functions. However if there is a significant proportion of
younger stars, the true selection function would be higher than the one we
used at the metal-richer and shorter-distance end. Using our selection
function will lead to an EDF that over-estimates the number of metal-poor
and distant stars.  The ability of our EDF to reproduce the density profile for K
giants found by \cite{xue+15} suggests  that our selection function is not
materially in error.
\subsubsection[]{Parameterised form of the EDF}
The imposition of a functional form on the EDF can bias the results by
restricting the set of possible solutions.  We believe the phase-space
distribution function is sufficiently general, allowing a range of density
and anisotropy profiles, but the dependence of the EDF on metallicity is
rather limited in that only the power-law indices $\alpha$ and $\beta$ depend
on [Fe/H], not the scale action $J_0$ or the parameters $a_i$
and $b_i$ that determine the shape of the velocity ellipsoid and the halo's
flattening. It is clear however that a metallicity gradient is not forced by
the EDF; if none were needed by the data,  the indices would have been found
to be
independent of metallicity.
\subsubsection[]{Adoption of a fixed potential} 
With sufficiently perfect data for a stellar population, one should be able
to determine the potential, because an EDF can perfectly model the data only
in the true potential. We chose a specific potential at the outset and sought
an EDF that reproduced the observations in this potential, and imperfections
in our choice of potential will both bias our EDF away from the true EDF and
give rise to discrepancies between our best model and the data. Our success
in reproducing the data suggests that our chosen potential is not seriously
in error.  The main weakness of our work probably lies in the small number of
very distant stars in the sample and the large uncertainties in their
tangential velocities.

\section[]{Conclusions and further work}
We used a sample of I-color K giants from the SEGUE-II survey to probe the
chemodynamical structure of our galaxy's stellar halo. To these data we
fitted an EDF that provides for a density profile that gradually steepens
with radius from a shallow asymptotic slope at small radii to a steeper slope
at large radii. These slopes can vary with [Fe/H]. Using an EDF enables one
to build a dynamically consistent model that allows the phase-space
distribution of stars to vary with metallicity, whilst incorporating a
realistic selection function that takes into account restrictions on sky
positions, proper motions, apparent magnitudes, and colours. We find that our models
reproduce the observations well, presumably because the data for K giants are
not yet rich enough to resolve most halo substructures.

We find that the power-law slope of the density profile of the K giants
steepens from $\sim -2$ to $\sim -4$ in the outer halo. The halo is flattened
to axis ratio $q\simeq0.7$ on the inside but becomes almost spherical on the
outside. The overall metallicity DF is best captured with two lognormal
distributions, peaking at $\FeH\sim-1.5$ and $\FeH\sim-2.3$. The
metal-rich stars are more tightly confined in action space than the
metal-poor stars, with the consequence that the halo has a weak metallicity
gradient. The dispersion in metallicity at any point increases outwards.  The
metal-rich stars form a more flattened body than the metal-poor ones. The
extent of velocity anisotropy depends on whether the Sagittarius stream is
included or not. With the stream included, the velocity ellipsoid is
everywhere moderately radially biased, while when it is excluded, moderate
radial bias in the inner halo gives way to isotropy at large radii. 

There are several possible directions for further work. The
EDF could be applied to detect substructures in a denser sample of K giants
in the phase-space-metallicity domain, thus complementing the work of
\cite{janesh+16}, who look for substructures in SEGUE K giants using sky
positions, heliocentric distance, and radial velocity. The EDF could be
elaborated to include rotation and to allow for explicit dependence on [Fe/H]
of the parameters $a_i$, $b_i$ that jointly control the system's flattening
and velocity anisotropy. The EDF could be changed to make the transition
between the inner and outer asymptotic slopes of the density profile sharper.
The EDF could be elaborated to include dependence on $\aFe$.  The EDF could
be fitted to a larger sample of K giants and to other halo tracers, such as
blue horizontal branch stars -- we hope to report on these exercises shortly.

\section*{Acknowledgments}
The research leading to these results has received funding from the European Research
Council under the European Union's Seventh Framework Programme (FP7/2007-2013)/ERC
grant agreement no.\ 321067. The work was also supported through
grant ST/K00106X/1 by the UK's Science and Technology Facilities Council.

We thank GitHub for providing free private repositories for educational use,
thus enabling seamless version control and collaboration. PD thanks Jason
Sanders, Wyn Evans, Gus Williams, and Eugene Vasilyev for useful
conversations, and Jason Sanders, Paul McMillan and Til Piffl for
routines. Finally, PD would like to thank Xiang-Xiang Xue for providing the K
giant catalogues and completeness fractions for the SEGUE-II plates.

\bibliographystyle{mnras}
\bibliography{biblio}

\appendix

\label{lastpage}

\end{document}